\pdfoutput=1

\documentclass[a4paper, longbibliography]{article}
\usepackage[margin=1in]{geometry} % full-width
\usepackage{amsmath}
\usepackage{amsthm}
\usepackage{amssymb}
\usepackage{braket}
\usepackage[utf8]{inputenc}
\usepackage{color,soul}
\usepackage{hyperref}
\usepackage{enumitem} 
\usepackage{caption}
\usepackage{mathtools}
\DeclarePairedDelimiter{\ceil}{\lceil}{\rceil}
\usepackage{subcaption}
\usepackage[sort&compress,numbers,square]{natbib}
\theoremstyle{plain}

\theoremstyle{definition}

\usepackage{graphicx, color}
\graphicspath{{fig/}}

\usepackage{algorithm, algpseudocode} % use algorithm and algorithmicx for typesetting algorithms
\usepackage{mathrsfs} % for \mathscr command

\usepackage{lipsum}

\title{Quantum Computing Simulation of a Mixed Spin-Boson Hamiltonian and Its Performance for a Cavity Quantum Electrodynamics Problem }

\author{Maria Tudorovskaya$^{1*}$ \and David Mu\~{n}oz Ramo $^1$ }

\date{
	$^1$Quantinuum Ltd., 13–15 Hills Road, Cambridge, CB2 1NL, United Kingdom \\%
	$^*$Corresponding author: \tt{maria.tudorovskaya@quantinuum.com}\\
 October 17, 2023
}

\begin{document}
\maketitle

\begin{abstract}
In this paper, we aim to broaden the spectrum of possible applications of quantum computers and use their capabilities to investigate effects in cavity quantum electrodynamics (``cavity QED''). 
Interesting application examples are material properties, multiphoton effects such as superradiance, systems with strong field-matter coupling, and others.
For QED applications, experimental studies are challenging, and classical simulations are often expensive. Therefore, exploring the capabilities of quantum computers is of interest. Below we present a methodology for simulating a phase transition in a pair of coupled cavities that permit photon hopping. We map the spin and boson systems to separate parts of the register and use first-order Trotterization to time-propagate the wavefunction. The order parameter, the observable for the phase transition, is calculated by measuring the number operator and its square. We introduce a boson-to-qubit mapping to facilitate a multi-photon, multi-atom case study. Our mapping scheme is based on 
the inverse Holstein-Primakoff transformation. In the multi-photon regime, boson operators are expressed via higher-spin operators which are subsequently mapped on a circuit using Pauli operators. We use a Newton series expansion to enable rigorous treatment of the square root operator. We reproduce the results of classical simulations of a phase transition with a noiseless 6-qubit simulation. We find that the simulation can be performed with a modest amount of quantum resources. Finally, we perform simulations on noisy emulators and find that mitigation techniques are essential to distinguish signal from noise. 
\begin{description}
\item[Keywords:]
Quantum Computing, Cavity Quantum Electrodynamics, Boson Mapping
\end{description}
\end{abstract}

%\keywords{Suggested keywords}%Use showkeys class option if keywothe rd
                              %display desired
\maketitle

\tableofcontents

\section{Introduction}
\label{sec:intro}
	
In recent years, the quantum computing community has been focusing on developing efficient algorithms that would allow the study of energetics and dynamics of different physical systems, such as molecules or solids, which can be described by fermionic or spin Hamiltonians. A wide array of software is being developed to tackle this problem, each aiming to minimize the number of qubits and the circuit depth.

In this work, we study the dynamics driven by a Hamiltonian which contains bosonic and spin operators. Our interest is twofold. On one hand, we address the problem of efficient mapping of bosonic particles onto quantum circuits. On the other hand, this opens the door to modeling many more systems of various nature, including \color{black}time-dependent
\color{black} collective effects. 

Relevant literature includes classical studies of effects occurring in systems involving both bosonic and spin or fermionic particles, as well as quantum simulations of such mixed systems. Systems with strong field-matter coupling   \cite{Figueroa2018} and phase transitions   \cite{Zueco2019} are actively studied by the cavity quantum electrodynamics (or ``cavity QED'') community. An example of a mixed system relevant to quantum chemistry is molecular vibrations and
molecules coupled to an electric field  \cite{CalvoIbar2021}.  
Several studies dedicated to simulating systems involving bosons have been carried out in recent years covering a broad range of aspects. Open quantum systems with Markovian and non-Markovian dynamics are considered in    \cite{3mdpiDigitalQuantum, Head-Marsden2021}.
  Reference  \cite{1li2023efficient} introduces a variational basis state encoding algorithm for electron-phonon systems. 
Qudit quantum operators are considered in   \cite{2natureResourceefficientDigital}.
 A variational quantum algorithm for computing vibrational states on a molecule is presented in   \cite{6majland2021resourceefficient}. Dynamics on near-term quantum hardware for first-quantized systems is analyzed in   \cite{4wang_krstic_2022} while   \cite{5arxivQuantumCircuit} considers mapping bosons to fermions and then to qubits. An overview of near- and long-term approaches to bosonic systems, specifically on vibrational spectroscopy, is given in   \cite{7PhysRevA.104.062419}. The performance of dynamics simulations is discussed in \cite{8Jaderberg_2022, peng2023quantum, 9PhysRevResearch.3.043212}. We note that this literature overview is not exhaustive.

Let us now consider a Hamiltonian for a system involving different kinds of particles. Such a Hamiltonian would typically contain terms that count the number of particles (or excitations) for each category and an interaction term that mixes operators of different types. For example, consider mixing spin and boson particles, 
\begin{align}
\label{eq:H_mixed}
    H^{\text{mixed}} \equiv H_{\text{spin}} + H_{\text{boson}} + H_{\text{interaction}}
\end{align}

An example of such a system is the Rabi Hamiltonian for an atom interacting with an AC electric field. Here, the atom as a two-level system is represented by spin operators and the field is quantized and described by boson operators:
\begin{align}
\label{eq:Rabi}
    H^{\text{Rabi}} = \omega b^\dagger b +  \frac{\omega_0}{4} \sigma_{+} \sigma_{-} + g\sigma_x (b + b^\dagger), 
\end{align}

\noindent where $b^\dagger, \: b$ are bosonic creation/annihilation operators, and
$\sigma_{\pm} = \sigma_x \pm i \sigma_y$ are ladder operators. Note the absence of the factor $1/2$ in this definition. 
For spin 1/2, the spin operator is
\begin{align}
    \boldsymbol{S^{(1/2)}} = \frac{1}{2}\boldsymbol{\sigma},
\end{align}
and the elements of the $\boldsymbol{\sigma}$ vector are the Pauli matrices {$\sigma_x$, $\sigma_y$, $\sigma_z$}:
\begin{align}
\label{eq:paulis}
  \sigma_x = 
    \begin{Vmatrix}
    0 & 1 \\
    1 & 0
    \end{Vmatrix}, \:\:\: 
    \sigma_y = 
    \begin{Vmatrix}
    0 & -i \\
    i & 0
    \end{Vmatrix}, \:\:\:
    \sigma_z = 
    \begin{Vmatrix}
    1 & 0 \\
    0 & -1
    \end{Vmatrix}  
\end{align}

Here, $\omega$ and $\omega_0$ 
are the field frequency and the energy difference between the ground and excited states of the atom and $g$ is the interaction strength.

Assuming the frequencies are nearly in resonance, i.e. $|\omega - \omega_0| \ll \omega $, 
the rotating wave approximation, where the counter-rotating terms are neglected, is applied and leads to the so-called Jaynes-Cummings Hamiltonian: 
\begin{align}
\label{eq:Jaynes-Cummings}
    H^{\text{JC}} = \omega b^\dagger b +  \frac{\omega_0}{4} \sigma_{+} \sigma_{-} + \frac{g}{2}(\sigma_{+} b + \sigma_{-} b^\dagger), 
\end{align}

\noindent where ``JC'' stands for ``Jaynes-Cummings''. Note that although in this study we will continue to talk about the Jaynes-Cummings Hamiltonian, the discussed approach is general and applies to Hamiltonians that include other types of bosonic excitations. Additionally, this approach can include fermionic particles, such as electrons in molecules.

The JC and Rabi Hamiltonians  (Eqs. \ref{eq:H_mixed}, \ref{eq:Rabi}, \ref{eq:Jaynes-Cummings}) contain both spin and bosonic components, and these components have to be encoded into the qubit register for simulations on a quantum computer.
Several boson mappings are known and used in the literature, and the question of the efficiency of these encoding schemes is important in the context of the overall efficiency of quantum algorithms.  In this paper, we consider an approach in which spins and bosons are mapped to different parts of the qubit register. This approach builds upon previous work for one cavity - one photon systems considered in \cite{Fitzpatrick2021} which is based on the Holstein-Primakoff transformation. 
% However, we note that alternative approaches must be considered in the future. 

In addition to the mapping problem, one has to consider the number of measurements (or shots) required to simulate the physical property of interest on a quantum computer. This problem is particularly prominent in quantum chemistry simulations, where the number of shots required to obtain energies with chemical accuracy is prohibitive with near-term hardware. However, in cavity QED interesting phenomena often have a qualitative nature, such as, for example, does a certain phase transition occur? Thus, we expect that a modest number of shots may be sufficient to simulate these systems on a quantum computer. We aim to test this assumption with a simulation of a system with multiple numbers of cavities and photons and introduce the system below.

The paper is organized as follows. First, in Section \ref{sec:mapping}, we present various ways to map bosonic operators present in mixed Hamiltonians, including the Holstein-Primakoff transformation for higher spins. For the latter, we briefly discuss its efficiency as compared to other mappings. Due to the novelty of the proposed mapping scheme, it is essential to test it for a physical problem which, on one hand, is useful for real-world applications, and, on the other hand, has been well described classically so that the result can be compared exactly. We refer to the TKET compiler described in  \cite{Sivarajah_2021}.
In Section \ref{sec:qed}, we introduce the concept of a phase transition in coupled cavities and explain its significance. We then apply our method to the model. We show how to prepare the initial wavefunction to correspond to a Mott insulator and run the dynamics to demonstrate what Hamiltonian parameters enable the phase transition, and when the wavefunction of the system describes it as superfluid.
In Section \ref{section:noisy}, we present the results of simulations on noisy backend emulators. Finally, in Section \ref{sec:conclusions}, we proceed to the conclusions giving the outlook of future improvements.
In this paper, we use ``bosons'' and ``photons'' interchangeably.

\section{Boson mapping}
\label{sec:mapping}

\subsection{Some standard mapping schemes}
\label{sec:standard_mapping}

There are several ways to map bosonic states and operators to qubits \cite{Chin2022, Morley-short2015, Huang2021}. Standard mapping schemes include but are not limited to, direct one-to-one mapping or binary mapping. 
Within the most straightforward scheme,  one-to-one mapping, the size of the Fock space corresponds to the number of qubits available for boson encoding. $N+1$ qubits can be used to encode $N+1$ Fock states with up to $N$ bosons. Consider one mode denoted as $\chi$,
and spins up and down denoted as $\uparrow$ and $\downarrow$, respectively (see details in   \cite{Huang2021}). The states and the creation operator are mapped as follows:

\begin{align}
\ket{0}_\chi \:\: \leftrightarrow &\:\:\ket{ \uparrow_0\downarrow_1\downarrow_2............\downarrow_N } \\
\ket{m}_\chi\:\: \leftrightarrow &\:\:\ket{ \downarrow_0\downarrow_1\downarrow_2...\uparrow_m...\downarrow_N } \\
b^{\dagger}_\chi \:\: = & \:\:\sum_i \sqrt{i+1}\sigma_-^i\sigma_+^{i+1}
\end{align}

% This can be compared to the binary mapping which allows to mapping 2$^{N}$ states with N qubits. For example, let us take a look at the matrix representation of the annihilation operator in the oscillator basis:

Binary mapping is more efficient in terms of qubit usage:
\begin{align}
\ket{0}_\chi \:\: \leftrightarrow &\:\: \ket{ \uparrow_1\uparrow_2...\uparrow_{t-1}\uparrow_t } \\
\ket{1}_\chi \:\: \leftrightarrow &\:\: \ket{ \uparrow_1\uparrow_2...\uparrow_{t-1}\downarrow_t } \\
\ket{2}_\chi \:\: \leftrightarrow &\:\: \ket{ \uparrow_1\uparrow_2...\downarrow_{t-1}\uparrow_t } \\
\ket{3}_\chi \:\: \leftrightarrow &\:\: \ket{ \uparrow_1\uparrow_2...\downarrow_{t-1}\downarrow_t } \\
\ket{2^t-1}_\chi\:\: \leftrightarrow &\:\: \ket{ \downarrow_1\downarrow_2...\downarrow_{t-1}\downarrow_t } 
\end{align}

Note that the register size $t \le N$. 

Once the states are mapped, one can proceed with either deriving the qubit representation of the operators or mapping the entire Hamiltonian on the circuit. 
 Although matrices describing bosonic operators have infinite dimensions, practical applications require truncation. An $N\times N$ matrix can represent boson creation, annihilation, and number operators, as well as the bosonic part of the Hamiltonian operator, for a system of up to $N-1$ photons. 
The number operator matrix has numbers from 0 to $N-1$ on its diagonal. 
Whether the numbers are ascending or descending is a matter of basis states ordering, but care needs to be taken that the creation and the annihilation operators are constructed consistently. 
% By definition, the annihilation operator matrix elements are given by $ \braket{ n|b|n'} = \sqrt{n'} \delta_{n,n'-1}$.

There is a variety of approaches to circuit mapping of bosonic operators. For example, in \cite{Huang2021} the form of the creation operators is derived separately and iteratively for different qubit numbers. The authors reason about the effect of the operator on the qubit register and use recurrence relations. For 2 qubits:
\begin{align}
\label{eq:bdagger_binary}
b^{\dagger}_\chi =  
    \frac{1}{4}(I + \sigma_z)\otimes\sigma_- +
    \frac{\sqrt{2}}{4}\sigma_-\otimes\sigma_+ +
    \frac{\sqrt{3}}{4}(I - \sigma_z)\otimes\sigma_-
\end{align}

\noindent where $I$ is a one-qubit identity operator,  
 and $\sigma_{\pm} = \sigma_x \pm i \sigma_y$ (see Eq. \ref{eq:paulis} for definitions). Note that, in this case, $\sigma_{\pm}$ differs from the one in   \cite{Huang2021} by the 1/2 coefficient, and therefore Eq. \ref{eq:bdagger_binary} also differs from   \cite{Huang2021}. Its matrix form is

\begin{align}
\label{eq:bdagger_matrix_huang}
b^{\dagger}_\chi = 
    \begin{Vmatrix}
    0 & 0 & 0 & 0 \\
    1 & 0 & 0 & 0 \\
    0 & \sqrt{2} & 0 & 0 \\
    0 & 0 & \sqrt{3} & 0
    \end{Vmatrix} .
\end{align}
Other bosonic operators can be derived correspondingly.
As mentioned, another approach to implementing the binary mapping starts with expressing the entire Hamiltonian as a Hermitian $2^n \times 2^n$ matrix. 
The matrix of the evolution operator
$e^{-iHt}$ is a unitary that can always be represented as a sequence of Pauli gates. The Hamiltonian itself can be represented as a linear combination of n-fold tensor products of the Pauli matrices. While it can always be done, this approach is inefficient, especially for a large number of bosons.
The question of what exactly is the most efficient way of representing this matrix on a circuit remains open.

\subsection{Holstein-Primakoff inverse mapping for multiphoton regime}
\label{sec:holstein-primakoff}

In this paper, we propose a scheme that borrows ideas from the higher spin (multiphoton) Holstein-Primakoff mapping. This physically motivated mapping scheme allows for treating spin and boson operators on an equal footing, 
\color{black} that is, in a mixed spin-boson system, one may choose to express the entire Hamiltonian either with spin operators or with boson operators. \color{black}

The direct Holstein-Primakoff transformation was introduced to treat high spin operators in ferromagnetic materials \cite{PhysRev.58.1098}. It converts spin particle operators in a mixed-particle Hamiltonian $H$ into bosonic ones,
\begin{align}
    & H(b, b^{\dagger}, S_{z, \pm}) \rightarrow H(b, b^{\dagger}, \Tilde{b}, \Tilde{b^{\dagger}}) \text{ \:\:\:direct} 
\end{align}    
Here, $b^{(\dagger)}$, $\Tilde{b}^{(\dagger)}$ symbols are for boson operators and $S{_z, \pm}$, $\Tilde{S}{_z, \pm}$ symbols are the ladder operators for the general spin operator $\boldsymbol{S}$ not limited to spin 1/2.

The inverse transformation converts bosonic operators into spin operators which is convenient for the quantum circuit model of quantum computing.
\begin{align}
    & H(b, b^{\dagger}, S_{z, \pm}) \rightarrow H(S_{z, \pm}, \Tilde{S}_{z, \pm}) \text{ \:\: inverse}
\end{align}
 Note that the spin operators may describe high-value spins. 

 \color{black} Different device architectures may perform differently for the same physical system. Performance also depends on the chosen mapping scheme. For a qubit/qudit architecture, expressing the Hamiltonian with spin operators is a natural choice. Besides physical intuition, the possible advantage of using the spin representation comes from the matrix form of the operators. Both bosonic and spin ladder operators are represented by sub- or superdiagonal matrices. In addition, higher spin matrices possess mirror symmetry relative to the main diagonal. Some studies suggest that such symmetry can lead to circuit depth reduction that includes the block-encoding of the non-unitary operator itself \cite{huang2023optimized}, \cite{Sunderhauf2024blockencoding}.
 
\color{black}So far, inverse mapping has received less attention, especially in the context of quantum computing, and is the subject of this paper.

Using the inverse Holstein-Primakoff transformation involves two steps: (i) expressing \color{black} each bosonic ladder operator
\color{black}in terms of higher-spin operators which can be done exactly or approximately, and (ii) mapping higher-spin operators to the quantum computer. For (ii), there is freedom in how exactly the mapping is performed, which depends on the system's highest spin. This, in turn, corresponds to the number of bosonic excitations one aims to represent, and whether a qubit, qutrit, or qudit architecture is used. So far, to our knowledge, the inverse Holstein-Primakoff mapping for quantum computers has only been used for spin 1/2 (see an example in   \cite{Fitzpatrick2021}), and it has not been considered for a broader range of physical multi-photon problems.

The classical problem of ``spinorization'' of a boson was introduced in   \cite{Maria2018}. One can write

% \begin{align}
%     &S_z = S\boldsymbol{I} - b^{\dagger}b  \\
% %    & b^{\dagger}b = SI \boldsymbol{+} S_z \\
%     & S_- = \sqrt{2} h(b^{\dagger}b) b = \sqrt{2} h(SI \boldsymbol{+} S_z) b\\
%     & S_+ = b^{\dagger}\sqrt{2} h(b^{\dagger}b) = b^{\dagger}\sqrt{2} h(SI \boldsymbol{+} S_z) \\
% \end{align}

\begin{align}
    \label{eq:hp_ladder}
    & b^{\dagger} = S_+ \: \frac{1}{ \sqrt{ S \boldsymbol{I} - S_z}}; 
    \:\:\:\:\:\: b =  \frac{1}{ \sqrt{S \boldsymbol{I} - S_z}} \: S_-  
\\
    & b^{\dagger}b \equiv S \boldsymbol{I} + S_z \label{eq:boson_number}
\end{align}
Here, $S$ is the maximum eigenvalue of $S_z$, $\boldsymbol{I}$ is a multi-qubit identity. Eq. \ref{eq:hp_ladder}, \ref{eq:boson_number} apply to any $S$. It is easy to see using the example of $S=1/2$ that the number operator has descending integers on the diagonal which differs from the usual definition. It is interesting to note that already at this point one can see the hint to why the mapping is always possible. In the z-basis, the $S_z$ matrix is always diagonal, with the diagonal elements $S_z^{n,n} = S + 1 - n$ and, therefore, the matrix representation of the number operator in Eq. \ref{eq:boson_number} is exactly the one in the oscillator basis as described in Subsection \ref{sec:standard_mapping}.

The next obstacle to overcome is the question of how to deal with the inverse square root operator. Combining Eq. \ref{eq:hp_ladder} and Eq. \ref{eq:boson_number},
\begin{align}
\sqrt{S \boldsymbol{I} - S_z} = 2S\sqrt{\boldsymbol{I} - 
\frac{b^\dagger b}{2S}}  
\end{align}
In the literature, a Taylor expansion has sometimes been employed (see, for example,   \cite{Fitzpatrick2021}). For a given function $f(b^{\dagger}b)$, its Taylor series expansion around zero reads:
\begin{align}
    f(b^{\dagger}b) = \sum_{k=0}^\infty \frac{1}{k!}\partial_{b^{\dagger}b}^k f(0) (b^{\dagger}b)^k
\end{align}
Strictly speaking, however, the Taylor expansion is only valid when the function is analytic around the expansion point which does not hold for the square root function. 

Following   \cite{Konig2021}, we employ the Newton expansion. This expansion allows us to represent the square root as a power series of the number operator (up to the power of 2$S$). 
Expressing the normalized matrix square-root function, $h$, in terms of the number operator, from   \cite{Maria2018},
\begin{align}
\label{eq:hp_expansion}
     h(b^{\dagger}b) = \sqrt{\textbf{I} - \frac{b^{\dagger}b}{2S}} = 
 \sum_{k=0}^{2S} \frac{b^{\dagger k} b^k}{k!}  \sum_{l=0}^k (-1)^{k-l} \sqrt{1-\frac{l}{2S}} {k \choose l}    
\end{align}

This expansion holds without assuming that (a) the total spin is large, and, therefore, the series can be truncated, or (b) that the small qubit register allows to capture a significant part of the essential physics for spin 1/2, (as in   \cite{Fitzpatrick2021}). The difference from the Taylor series can be seen in the expansion coefficients starting with the power $k=1$. For the Newton expansion, $  h \approx 1I - b^{\dagger}b $. Compare it to Taylor expansion: $ h \approx 1I - b^{\dagger}b/2 $. The advantages of the Newton series are that it is exact for $2S$ terms in the expansion and that the square root function is now expressed in terms of the number operator which, we know, is diagonal. Inverting a diagonal operator or finding powers of a diagonal operator is a trivial task. Quantum algorithms for working with diagonal operators are also well-known \cite{diagonal}.

In the system of interest we are discussing below, the maximum spin corresponds to $2S = 3$. In this case, we have
\begin{align}
\label{eq:h_function}
    h = \alpha \boldsymbol{I} + \beta b^{\dagger}b + \gamma (b^{\dagger}b)^2 + \delta (b^{\dagger}b)^3,
\end{align}

\noindent where $\alpha, \beta, \gamma$ are the expansion terms which can be found from Eq. \ref{eq:hp_expansion}. Eq. \ref{eq:h_function} describes a diagonal matrix, and, therefore finding the matrix for $h^{-1}$ when performing the inverse transformation is straightforward. 

\subsection{Circuit implementation of the Holstein-Primakoff mapping for spin 3/2} 
\label{subsec:hp_circuits}

We present here circuits for our model. We focus on the specific case corresponding to $S = 3/2$, which represents the maximum spin in the system under study that we discuss below.

We suggest the following way to map spin 3/2 operators onto qubit Pauli operators, which is exact:
\begin{align}
\label{eq:Sx}
    2S_x &=
    \sqrt{3}I \otimes \sigma_x + \frac{\sigma_- \otimes \sigma_+ + \sigma_+ \otimes \sigma_-}{2} \\
    \nonumber 
    &=
    \sqrt{3}I \otimes \sigma_x + \sigma_x \otimes \sigma_x + \sigma_y \otimes \sigma_y
\end{align}

\begin{align}
\label{eq:Sy}
    2S_y &= 
    \sqrt{3}I \otimes \sigma_y -  \frac{\sigma_- \otimes \sigma_+ - \sigma_+ \otimes \sigma_-}{2i} \\
    \nonumber
    &=
    \sqrt{3}I \otimes \sigma_y + \sigma_y \otimes \sigma_x - \sigma_x \otimes \sigma_y
\end{align}

\begin{align}
\label{eq:Sz}
    2S_z & 
  = \frac{(I + \sigma_z) \otimes (I - \sigma_z) - (I - \sigma_z) \otimes (I + \sigma_z)}{4} + \\
   &  + 3 \: \frac{(I + \sigma_z) \otimes (I + \sigma_z) - (I - \sigma_z) \otimes (I - \sigma_z)}{4}
     \nonumber
\end{align}

The ladder operators are $S_{\pm} = S_x \pm iS_y$, as usual.

Using Eq. \ref{eq:hp_ladder} - Eq. \ref{eq:Sz}, we obtain
\begin{align}
\label{eq:bdagger_binary_reordered}
    b^{\dagger} = 
     \frac{\sqrt{1}}{4}(I - \sigma_z)\otimes\sigma_+ 
     + 
    \frac{\sqrt{2}}{4}\sigma_+\otimes\sigma_- 
     + 
    \frac{\sqrt{3}}{4}(I + \sigma_z)\otimes\sigma_+
\end{align}

The expression in Eq. \ref{eq:bdagger_binary_reordered} can be obtained from Eq. \ref{eq:bdagger_binary}
by multiplying with  $\sigma_x \otimes \sigma_x$ both on the left and on the right. The matrix representation of Eq. \ref{eq:bdagger_binary_reordered} is

\begin{align}
\label{eq:bdagger_matrix}
   & b^{\dagger} = 
    \begin{Vmatrix}
    0 & \sqrt{3} & 0 & 0 \\
    0 & 0 & \sqrt{2} & 0 \\
    0 & 0 & 0 & \sqrt{1} \\
    0 & 0 & 0 & 0
    \end{Vmatrix} 
\end{align}

The difference in the order of the elements between Eq. \ref{eq:bdagger_matrix} and  Eq. \ref{eq:bdagger_matrix_huang} corresponds to the different (ascending or descending) order of the matrix elements in the number operator and is a matter of choice. In this work, to make our matrix representation of the Hamiltonian exactly match the classical one from \cite{Norambuena2020}, we further multiply the Hamiltonian operator on the left and the right by the string of $\sigma_x$ operators acting on all qubits. The same effect, however, can be achieved by altering the meaning of qubits and their states for each cell.

We stress that although Eq. \ref{eq:bdagger_binary} and Eq. \ref{eq:bdagger_binary_reordered} are equivalent, the Holstein-Primakoff transformation does not require iterative operator construction. Its circuit mapping relies on the mapping of the higher spin matrices - which are known and standard -  and the diagonal matrices' circuit representation.

% % \noindent The creation operator looks similar except the non-zero values are below the diagonal. The number operator $b^\dagger b$ is a diagonal matrix with numbers from 0 to N-1 on the diagonal.

\subsection{Efficiency of the mapping scheme}
\label{subsec: efficiency}

 In this section, we assess the efficiency of the mapping scheme for time evolution. While efficient decomposition techniques for exact unitaries exist, they are normally limited to a small number of qubits making it essential to analyze the approximate evolution. 
 
 Recalling the exponential form of the time evolution operator for the full Hamiltonian Eq. \ref{eq:H_mixed} and applying the Trotter-Suzuki expansion, 
  \begin{align}
 \label{eq:evolution}
     U_{H_{\text{mixed}}}(dt) &\approx e^{-iH_{\text{mixed}} dt} \approx\\
     &\approx e^{-iH_{\text{spin}} dt} e^{-iH_{\text{boson}} dt} e^{-iH_{\text{interaction}} dt}
     \nonumber
 \end{align}
 
 Below, we carry out an empirical benchmarking analysis for the exponential operator in Eq. \ref{eq:evolution} in the form of Eq. \ref{eq:Rabi} representing the interaction.  We chose this term since it contains both spin and boson operators.  In this analysis, we consider a minimal example of a multi-boson coupled system, namely, a single atom (spin) interacting with a multi-photon field (i.e. enabling multiple bosonic excitations). First, we look closely at the case with up to 3 photons. After that, we show how it scales with the number of photons.

 \subsubsection{Up-to-3 photons case}
The maximum of 3 photons corresponds to the Holstein-Primakoff mapping with $S = 3/2$ as in Eq. \ref{eq:Sx} - \ref{eq:Sz}.

Here, we compare the circuits described in (a) and (b):
 
 \noindent (a) an efficient circuit for the exact evolution operator
 \begin{align}
\label{eq:exp_circ}
   U &= e^{-iH_{\text{interaction}} \tau} =\\
     &=  e^{- i g \sigma_x(b + b^\dagger) \tau} |_{(\tau = 1, g = 1)} = e^{- i \sigma_x(b + b^\dagger) }.
     \nonumber
\end{align}
\color{black} We begin by obtaining the matrix representation of 
$U$, which is derived as the matrix exponential of the tensor product of $-i\sigma_x$ and $b+b^{\dagger}$; 

 \color{black}
 \noindent(b) an approximation, $V$, mapped to a circuit using the multiphoton Holstein-Primakoff transformation.  
 
\begin{itemize}[leftmargin=0. in]
\item[]   
\begin{center}
    (a)
\end{center} 

Consider the evolution operator in Eq. \ref{eq:exp_circ}. The spin operator and the boson operators are mapped on separate qubits. Each boson operator is to be eventually expressed in the form of linear combinations of Pauli strings which are not generally commuting, therefore the Trotter-Suzuki decomposition is not exact. 
The matrix representation of $U$, in the case of a maximum number of 3 bosons is:

\begin{align}
U = 
     \begin{Vmatrix}
    0.023 & 0. & -0.714 & 0. & 0. & -0.632j & 0. & 0.301j   \\
  0. & -0.56 & 0. & -0.412 & -0.632j & 0. & -0.342j & 0.\\
 -0.714 & 0. & 0.023 & 0 & 0.  & -0.342j & 0. & -0.61j\\
 0. & -0.412 & 0. & 0.606 & 0.301j  & 0. & -0.61j & -0.\\
 0. & -0.632j & 0. & 0.301j & 0.023 & 0. & -0.714 & 0.\\
 -0.632j & 0. & -0.342j & 0. & 0. & -0.56 & 0. & -0.412 \\
 0.  & -0.342j & 0. & -0.61j & -0.714 & 0. & 0.023 & 0.\\
 0.301j & 0. & -0.61j & 0. & 0. & -0.412 & 0. & 0.606\\
 \end{Vmatrix} 
 \label{eq:Umatr}
\end{align}

The $8\times8$ matrix of $U$ is mapped to 3 qubits.
For 1, 2, and 3-qubit unitaries, TKET    \cite{Sivarajah_2021} can construct optimal (in terms of the CNOT count) circuits relying on the improvement to the Cosine-Sine Decomposition (CSD)  \cite{Tucci1, Shende_2004, Shende_2006}. For 3 qubits, the TKET  method is called ``Unitary3qBox''. The
Shannon decomposition used in the algorithm builds on the CSD decomposition and also utilizes the fundamental and widely-used KAK decomposition   \cite{Tucci} introduced by Helgason   \cite{Helgason} and implemented by Khaneja and Glaser   \cite{Khaneja2000} in one of the steps. 
The unitary matrix for Eq. \ref{eq:exp_circ} is then automatically converted to a 3-qubit circuit box. 
The circuit is subsequently optimized with the removal of possible redundancies introduced during the initial circuit construction process. 
\vspace{0.8cm}
\item[]
\begin{center}
    (b)
\end{center}

We construct an approximate circuit for the operator $U$ using the Trotter-Suzuki decomposition for Eq. \ref{eq:exp_circ}. {The Trotter-Suzuki decomposition for this Hamiltonian is not exact since $\sigma_x b$ does not commute with $\sigma_x b^{\dagger}$ in the exponential operator in Eq.\ref{eq:exp_circ}. When mapped on a circuit, the argument of the exponential operator will also contain non-commuting Pauli strings. Let $V$ be a unitary corresponding to the Trotter-Suzuki decomposition with \color{black}N$_{\text{trotter}}$ slices}:
\begin{align}
\color{black}
\label{eq:exp_circ_trot}
    % V = \prod^text{N}_{trotter} \left( e^{- i\sigma_x(b + b^\dagger)/text{N}_{trotter}} \right) \\
   V = \left( e^{-i\frac{\sigma_x b}{ \text{N}_\text{trotter}}} e^{-i\frac{\sigma_x b^{\dagger}} {\text{N}_\text{trotter}} }\right)^{\text{N}_\text{trotter}}
\end{align}
\color{black}
\noindent choosing N$_{\text{trotter}}$
\color{black}such that the matrix form of $V$ is a good approximation of $U$. 

To map each term in the product, we use the Holstein-Primakoff encoding as described in sections \ref{sec:holstein-primakoff}. and \ref{subsec:hp_circuits}. 
\noindent When N$_\text{trotter} = 10$, $ \left |\frac{V_{ij} - U_{ij}}{ \max{U_{ij}}  }\right | = \mathcal{O}(0.1)$, which is an acceptable Trotter error for this study.
The resulting matrix is given below in Eq. \ref{eq:Vmatr} to be compared to Eq. \ref{eq:Umatr}. 

\begin{align}
\label{eq:Vmatr}
&  V = 
 \begin{Vmatrix}
    0.021 & 0 & -0.668 & 0. &  0. & 
   -0.682j  &   0.  &  0.298j \\
 0. &  -0.556 &  0.  &-0.456 &  -0.581j &
   0. &   -0.382j & 0.   \\
 -0.757 & 0. & 0.023 &   0.  &   0.  &  -0.3j   & 0. &   -0.58j \\
  0. &   -0.37  &   0.    & 0.606 &   0.298j &  0.  &  -0.638j & 0.   \\
  0. &    -0.682j & 0.  &  0.298j & 0.021  & 0. &  -0.668  &   0.   \\
-0.581j & 0. &   -0.382j & 0.  &  0. & -0.556&   0. &   -0.456 \\
  0. &   -0.3j  &  0. &   -0.58j & -0.757& 0.  &   0.023&  0. \\
 0.298j & 0.  &  -0.638j & 0. &  0. & -0.37 &   0.  &  0.606  
 \end{Vmatrix} 
\end{align}

We compare circuits representing the unitaries (a) and (b) in terms of the number of native two-qubit operations when compiled with TKET  at the highest optimization level. The uncompiled circuit in (b) is significantly longer by construction (since we append N$_{\text{trotter}}$ copies of the same sub-circuit), and the aim is to show that with efficient compilation it contains no more 2-qubit gates than (a) while approximating the effect of the same unitary. The optimization
leads to removing redundancies, simplifies specific known sequences of Clifford gates, and makes use of circuit decomposition techniques.
\cite{Sivarajah_2021}. The circuit optimization is performed globally for each time step rather than for each Trotter step.  In the last step, the circuits are rewritten in terms of the machine's native gates, where the machine is either IBM's superconducting architecture or an ion trap architecture (Quantinuum H1 device).

\begin{table}[h!]
    \centering
    \begin{tabular}{|c|c|c|}
    \hline
  & IBM & H1  \\
      \hline
$U$ (TKET  Unitary3qBox)  & 24 (CX) & 19 (ZZ-phase) \\
\hline
$V$ (Holstein-Primakoff)  & 25 (CX) & 17 (ZZ-phase) \\
\hline
    \end{tabular}
    \caption{Number of 2-qubit gates in exact unitary $U$, Eq. \ref{eq:exp_circ} and its Trotterized approximation, $V$, Eq. \ref{eq:exp_circ_trot} circuit representation. The circuits are compiled for the corresponding gate sets native to the architecture, 
    i.e. superconducting (IBM) or trapped ion (H1). The type of the 2-qubit gates is given in brackets. The optimization level in TKET  is 2  \cite{Sivarajah_2021}.}
    \label{tab:benchmarking}
\end{table}

Table \ref{tab:benchmarking} shows how many native 2-qubit gates circuits for (a) and (b) contain. One can see that compiling for an H1 device leads to a slightly smaller 2-qubit gate count than compiling for an IBM machine. Moreover, assembling the circuit from individual spin and boson operators mapped using the Holstein-Primakoff transformation with subsequent compilation turns out to be slightly more efficient for H1 than mapping the matrix for the exact unitary $U$ directly onto the circuit. For a larger number of qubits, such analysis would have to be done separately, although we anticipate that this general conclusion holds.

\end{itemize}

By comparing the circuits in (a) and (b), we have shown that the Holstein-Primakoff mapping is not only intuitive and physically motivated but also efficient for mixed spin-boson Hamiltonians. 

\begin{figure}[h!]
    \centering
    \includegraphics[width=5.5cm]{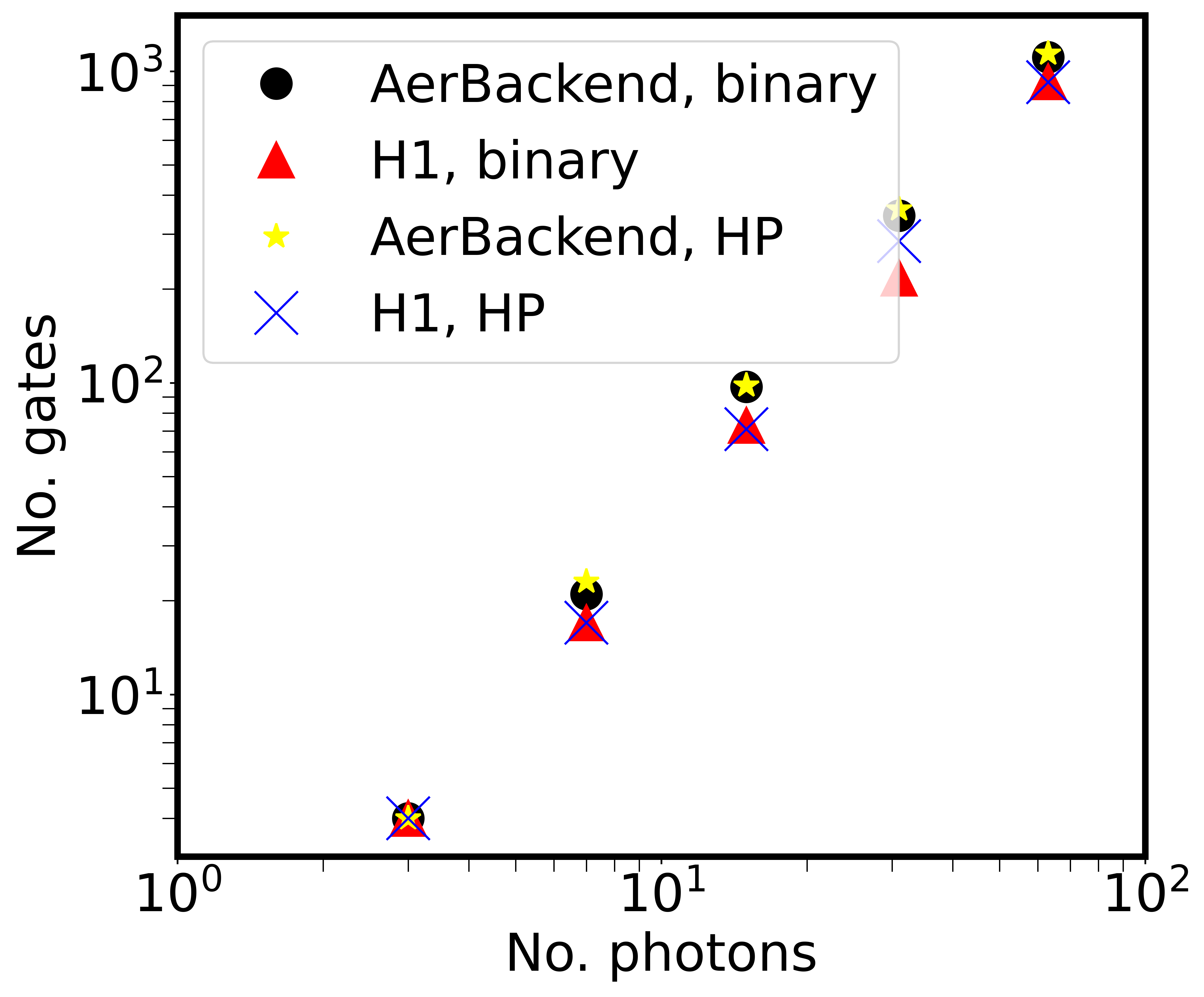}
    \caption{The number of 2-qubit gates in a circuit representation of the interaction term unitary, Eq. \ref{eq:evolution}}.
    \label{fig:photon_n_scaling}
\end{figure}

 \subsubsection{Higher number of photons}

Here we consider how the circuit depth for a unitary in Eq. \ref{eq:evolution} in terms of 2-qubit gates depends on the maximum number of photons allowed in the system.  Whether we map the $b, b^{\dagger}$ matrices of bosonic operators (Eq. \ref{eq:bdagger_matrix}) or use the higher-spin matrices (see Eq. \ref{eq:hp_ladder}, \ref{eq:boson_number}), we rely on the expansion as a linear combination of Pauli strings as follows:

\begin{align}
\label{eq:pauli_string}
\color{black}\sum_l C_lP_l = \sum_l C_l\prod_{\kappa=0}\limits^{\kappa_{\text{max}}} \sigma_{i_\kappa}^{(q_\kappa)},
\end{align}
where $C_l$ is the coefficient of the $l^{th}$ Pauli string $P_l$, $\sigma_i=\{\sigma_x,\sigma_y,\sigma_z\}$ determines the kind of Pauli gate, and $q$ is the qubit the corresponding gate acts on. While different representations can be valid, it is always possible to choose $C_l = \frac{1}{4}Tr(P_l \cdot U)$. For Holstein-Primakoff mapping, we find the qubit representations for the inverse square root operator and the spin ladder operators separately. Once the qubit representation is found, the operator is exponentiated using the Trotter-Suzuki decomposition, and the resulting circuit is optimized with TKET for the corresponding backend. We look at how deep the circuits are in terms of 2-qubit gates for spin mapping of the  Holstein-Primakoff transformation vs. binary mapping of the bosonic operators. 

The results are shown in Fig. \ref{fig:photon_n_scaling}. Upon the optimization, the two approaches lead to closely matching results. The plot exhibits linear dependency. Note that the axes have logarithmic scaling and that the number of qubits required is  $2^{\ceil{\log_2(\{ \text{No. photons}\}+1)}}$. 
The Qiskit AerBackend optimization leads to slightly deeper circuits. 

\color{black} It should be recognized that the circuit representation of the interaction term and the full evolution operator expressed with either spin or boson operators is not unique. As noted above, matrices of spin operators belong to the $SU(2)$ group and thus possess additional symmetry compared to boson operators. This implies that if a way of taking advantage of this symmetry is found, the efficiency compared to the binary encoding will extend to the multiphoton case. \color{black}

\begin{figure}[h!]
    \centering
    \includegraphics[width=8.5cm]{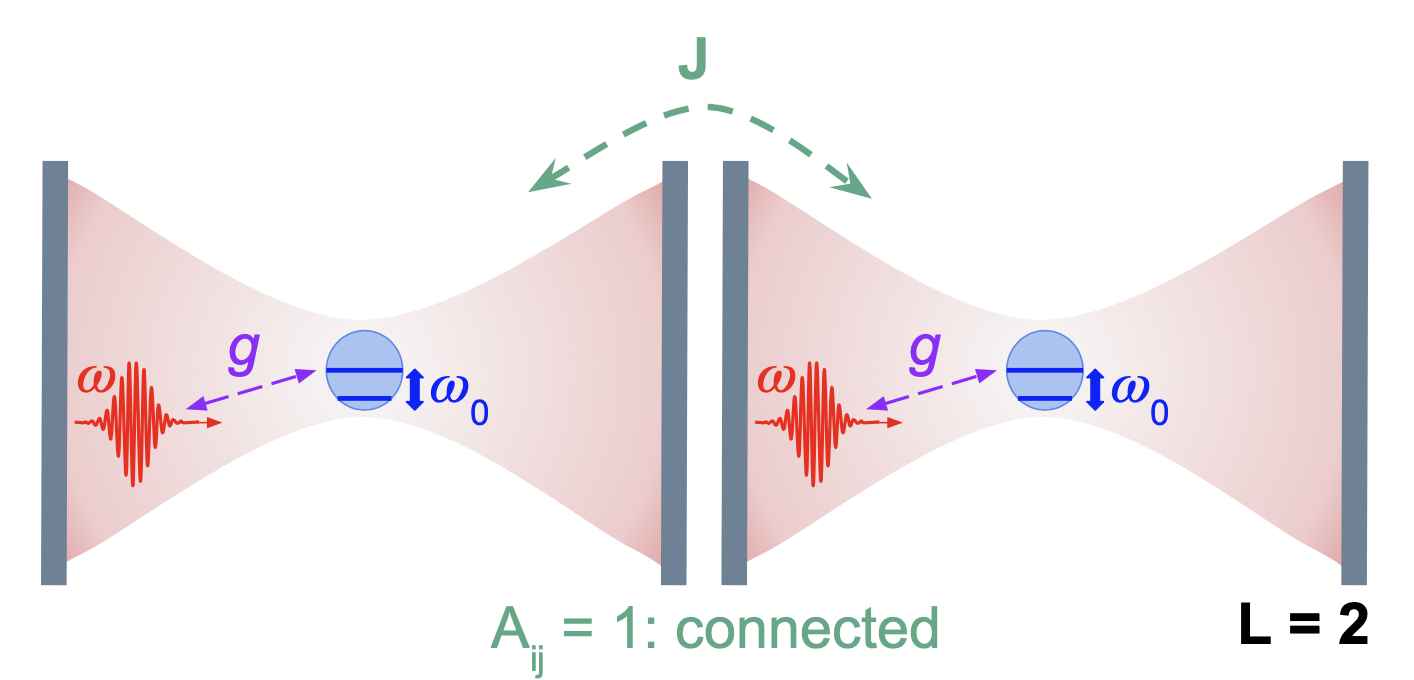}
    \caption{Schematic depiction of the two coupled cavities (L = 2, A$_{ij}$ = 1), each containing an atom (blue circle) and a mode of the electric field (red curve) with frequency $\omega$. The hopping strength is denoted by $J$ and the atom-light coupling is $g$. The energy gap between the ground and the excited state of the atom is $\omega_0$}.
    \label{fig:cavities}
\end{figure}

\section{Quantum electrodynamics in coupled cavities}
\label{sec:qed}

In the previous section, we selected a mapping scheme. In this section, we are applying it to a real-world problem. Our dual goal is: (a) to demonstrate that our approach is correct and can accurately replicate classical results in a statevector simulation, and (b) to show that QED problems mapped onto quantum circuits with the proposed scheme are suitable for performing a shot-based simulation with reasonable resources and extracting meaningful results.

Our example is the insulator-to-superfluid phase transition which is a purely many-body quantum 
phenomenon. In contrast to classical phase transitions, where a system undergoes a qualitative change in its macroscopic properties, a quantum phase transition is a result of quantum fluctuations that take place at temperatures close to absolute zero and can cause a sudden change in the system's quantum state. 

This effect was first proposed for liquid helium \cite{superfluid_fisher} and later discovered in other systems such as ultracold gases in optical lattices \cite{Greiner2002}.
Other examples are Josephson junction arrays, and antiferromagnets or frustrated spin systems in the presence of a magnetic field. 
In the superfluid phase, the particles are unbound and exhibit long-range coherence losing their individual character. In the Mott insulator phase, in contrast to the superfluid phase, the particles are confined and cannot conduct electricity or move freely.

Rapid progress in cavity and circuit electrodynamics made it possible to adjust and control the coupling strength between the spin and bosonic particles. Arrays of coupled cavities, where atoms are coupled to a mode of the electric field and photon hopping are allowed between the cavities, is therefore a convenient system to study the phase transition \cite{Houck_QED_2012, Figueroa2018}. The phase and the intensity of light can be controlled and physical observables can be measured in different regimes in real time. Our setup consists of two coupled cavities as shown schematically in Fig. \ref{fig:cavities} which is sufficient to observe the phenomenon \cite{Norambuena2020}.

\subsection{System Hamiltonian and parameters}
\label{subsec:system_params}

The physics of the system is well described by the so-called Jaynes-Cummings-Hubbard Hamiltonian, where the ``Hubbard'' part refers to the way the cavities interact, and each cavity is a Jaynes-Cummings system described in Eq. \ref{eq:Jaynes-Cummings} \footnote{Note that it is also possible to consider the Rabi-Hubbard Hamiltonian, which is, however, outside of the scope of this paper}. In the rotating wave approximation, the full Hamiltonian reads:
\begin{align}
\label{eq:H_JCH}
    H^{\text{JCH}} &= \sum^L_{i=1} H^{\text{JC}}_i - J\sum_{<i,j>}A_{ij}(b_i^\dagger b_j + b_i b_j^\dagger) = \\ \nonumber &= 
    \sum_{i=1}^L \big[\omega b^\dagger_i b_i +  \frac{\omega_0}{4} \sigma_{i,+} \sigma_{i,-} + \frac{g}{2}(\sigma_{i,+} b_i + \sigma_{i,-} b^\dagger_i) \big] - 
    \\ \nonumber
    & \hspace{1cm}- J\sum_{<i,j>}A_{ij}(b_i^\dagger b_j + b_i b_j^\dagger),
\end{align}

\noindent where the first summation goes over all the cavities, in our case up to $L=2$. JCH stands for ``Jaynes-Cummings-Hubbard'', and $J$ is the inter-cavity coupling strength. $A_{ij}$ is either 0 or 1 for each cavity pair depending on whether photon hopping is allowed or not between the cavities with the corresponding indices.  

Table \ref{tab:params} lists the parameters we use in the simulation. The characteristic time of the process is $T=1/J$. We choose to consider a system where $J$ is large, and as a result, the simulation period is shorter. The motivation for this choice is to maintain good accuracy using only a modest number of time steps and Trotter steps. Throughout the study, we vary $\Delta$ to cover both the near-resonant $\omega_0 - \omega \approx 0$ and off-resonant regimes.

\begin{table}[h!]
    \centering
    \begin{tabular}{|c|c|c|c|}
    \hline
      $\omega$ & $\Delta$  = $\omega_0$ - $\omega$     & $g$ & $J$  \\ \hline 
        1    & \{ 10$^{-5}g$ - 10$^{5}g$ \} & 0.1 & 0.1 \\
      \hline
    \end{tabular}
    \caption{Parameters of the 2 coupled arrays. The values of the parameters are given in atomic units.}
    \label{tab:params}
\end{table}

For each cavity, we utilize 1 qubit to represent the atom and 2 additional qubits for the photons. The maximum number of photons per cavity is determined by the initial conditions. In the setup, a cavity cannot accommodate more than 2 photons. However, the use of 2 qubits per cavity, in principle, allows us to include up to 3 photons per cavity. In general, the number of qubits required may always allow for a larger number of bosonic states beyond the specificity of the problem setup.

The resulting qubit Hamiltonian mapped to qubits contains 55 terms of the general form of Eq. \ref{eq:pauli_string}. The maximal length of a Pauli string in the considered qubit Hamiltonian is $\kappa_{\text{max}} = 4$.

\subsection{Initial state preparation}
\label{subsec:initial_state}

\begin{figure}[h!]
     
     \begin{subfigure}[b]{0.45\textwidth}
         \caption{}
         \centering
         \includegraphics[width=6.cm]{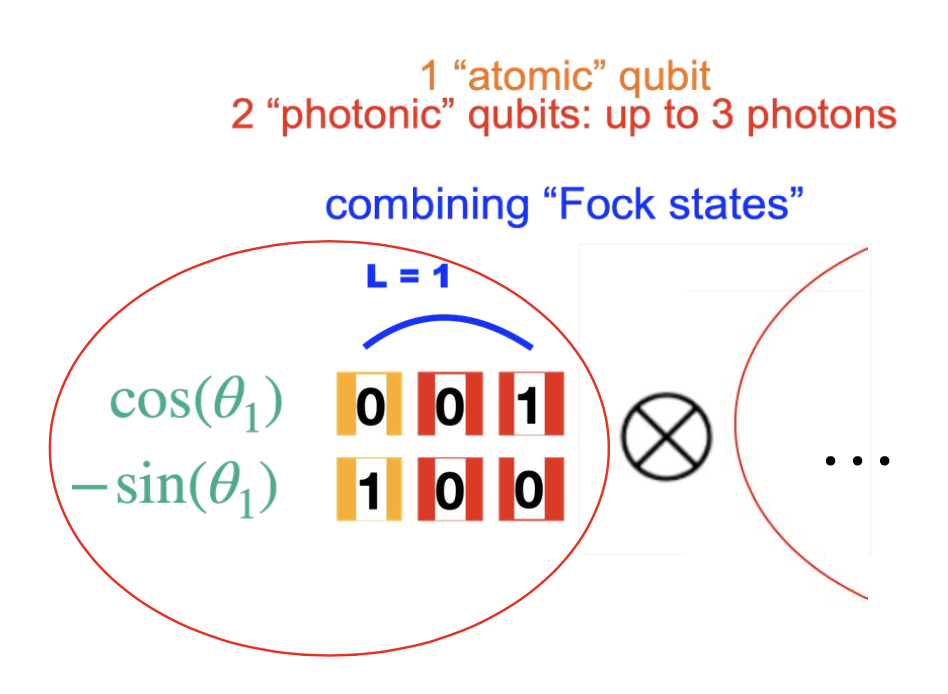}
         \label{fig:initial_state_qubit}
     \end{subfigure}
     \hfill
     \begin{subfigure}[b]{0.45\textwidth}
    \caption{}
         \includegraphics[width=7cm]{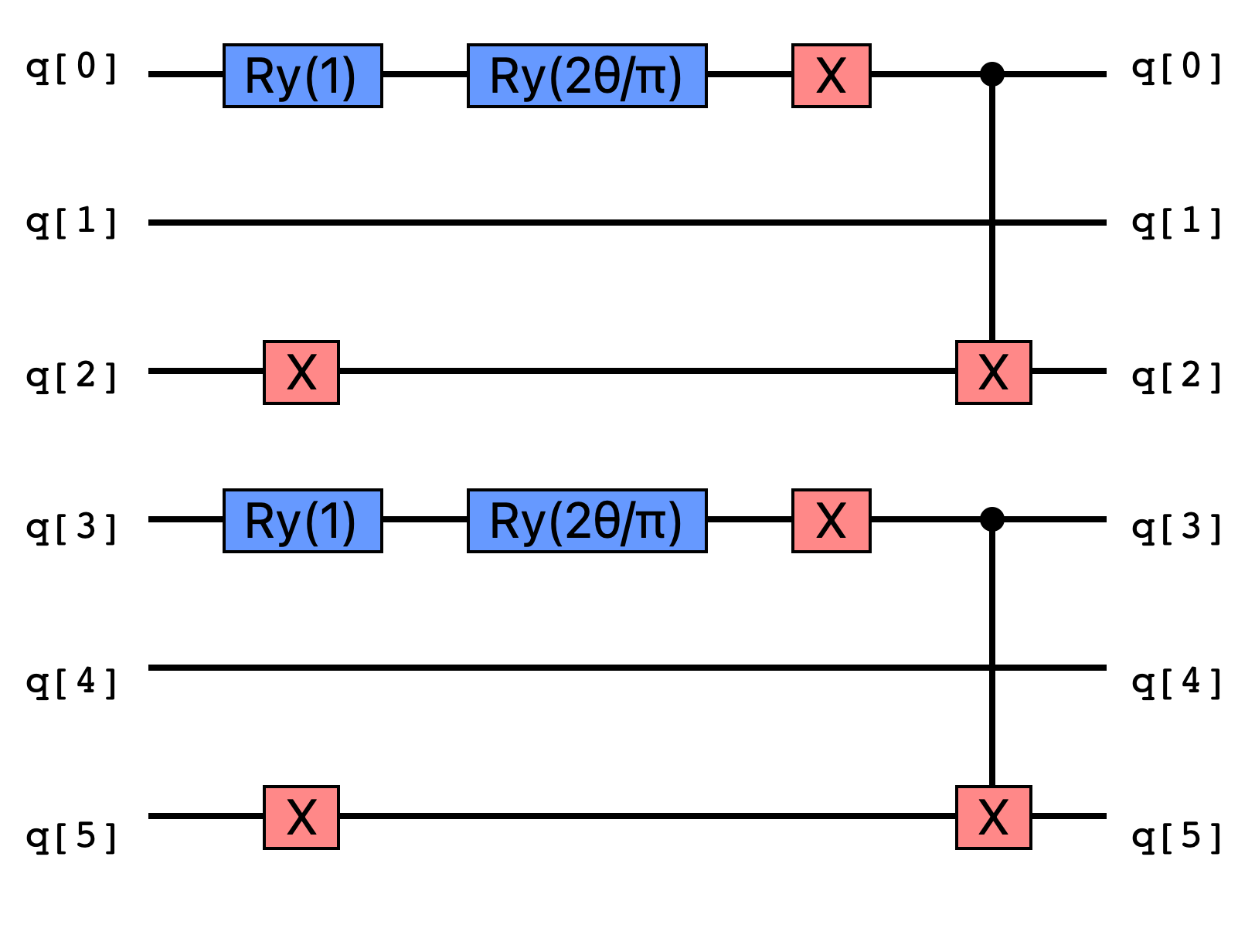}
         
        \label{fig:initial_state_circuit}
     \end{subfigure}

        \caption{
        Initial state corresponding to the Mott insulator phase. (a) Schematic representation of the initial state construction for one circuit box representing a single cavity. The initial state for a single cavity is a linear combination of 2 Fock states corresponding to either an excited atom and 0 photons, or an atom in the ground state and 1 photon. $\theta_1$ is a function of the detuning, $\Delta$. (b) Circuit for the state constructed using the Givens rotations ansatz and decomposed into rotations and Pauli gates. The angle of the rotation gate is a function of the detuning. Note the sub-circuits on the top 3 and the bottom 3 qubits of the register are identical.}
        \label{fig:three graphs}
\end{figure}

The initial state, denoted by $\Psi_{t=0}$, corresponds to a pair of identical cavities each in the Mott insulator state. In each cavity, the state $\ket{n=1,-}$, with $n$ being the number of photons, is a linear combination of Fock states. Only one type of excitation is allowed in this state, either an atom in the excited state or one photon present.
The resulting full initial state is constructed as a tensor product of $L$ individual cavity states   \cite{Figueroa2018, Norambuena2020}: 
\begin{align}
    \Psi_{t=0} = \ket{n,-}_0 \otimes...\otimes \ket{n,-}_L 
\end{align}
and
\begin{align}
    \ket{n,-}_i = \cos(\theta_i)\ket{e}\otimes\ket{n} - \sin(\theta_i) \ket{g}\otimes \ket{n-1} 
\label{eq:cavity_state}
\end{align}
In our setup, $L=2$ and the initial number of photons in a cavity $n=1$, $\theta_i$ is related to the coupling strength, $g$, the detuning $\Delta = \omega - \omega_0$ as
\begin{align}
    \tan(\theta_n) = 2g\sqrt{n}/\Delta
\end{align}
Specifically, we need to build an ansatz that combines the qubit states $\ket{100}$ and $\ket{001}$, as shown in Fig. \ref{fig:initial_state_qubit}. We use a particle-preserving ansatz that employs Givens rotations   \cite{Arrazola_2022}. Controlled Givens rotations are universal for particle-conserving unitaries, which allow us to explore the entire Hilbert space. The matrix representation of the Givens rotation by an angle $\phi$ is
\begin{align}
G = 
\begin{Vmatrix}
1   &... & 0 &  ... & 0 &  ... & 0 \\
.   & . & . &  . & . &  . & . \\
0   &... & c  & ... & -s &  ... & 0 \\
.   & . & . &  . & . &  . & . \\
0   &... & s &  ... & c &  ... & 0\\
.   & . & . &  . & . &  . & . \\
0   &... &  0 &  ... &  0 &  ... & 1     
\end{Vmatrix}
\label{eq:givens}
    \end{align}  
where $c=\cos{\phi}$ and $s=\sin{\phi}$.
% % It can be shown that a multiple-qubit multicontrolled general excitation can be decomposed into multicontrolled single excitations. 

Givens rotations with gates parametrized as in Eq. \ref{eq:givens} can be used to prepare a linear combination of Fock states with a fixed number of ones and zeroes on 3 qubits representing the first cavity. We then append an equivalent circuit box for the second cavity. The resulting circuit for the initial state (for a specific value of the detuning) is shown in Fig. \ref{fig:initial_state_circuit}.

\subsection{Time evolution of the wavefunction}

The dynamics of the system are described by the exponential evolution operator,
\begin{align}
    \Psi(t) = e^{-iH^{\text{JCH}}t} \Psi_0,
\end{align}

\noindent where $\Psi_0$ is the initial state described in Section \ref{subsec:initial_state} and $H^{\text{JCH}}$ is the Hamiltonian defined in Eq. \ref{eq:H_JCH}.

We rely on the Trotter-Suzuki transformation to propagate the wavefunction:
\begin{align}
\label{eq:Trotter}
    \Psi(t+dt) &= e^{-iH^{\text{JCH}}dt} \Psi(t) = \left( e^{-iH^{\text{JCH}}\Delta t} 
    \right)^{\color{black}\text{N}_{trotter}}\Psi(t) \approx
    \\\approx\nonumber
    & \prod\limits_l \left( e^{-iC_l P_l \Delta t} \right)^{\color{black}\text{N}_{trotter}}\Psi(t),
\end{align}

\noindent where $H^{\text{JCH}} = \sum\limits_l C_l P_l$ is a sum of Pauli strings $P_l$ with corresponding coefficients as in Eq. \ref{eq:pauli_string}, and \color{black}N$_{\text{trotter}} = dt/\Delta t$ 
\color{black}determines the number of Trotter steps per time step. 
We perform measurements at each time step up to the characteristic time $T=1/J$, where the value for the hopping strength $J$ is given in Table \ref{tab:params}.
Let us say we measure at time $t = m*dt$. First, we construct an elementary circuit corresponding to one Trotter step.  Then \color{black}N$_{\text{trotter}}*m$ 
\color{black}such circuits are appended to one another. Once this is done, we compile and optimize the resulting circuit, assuming no device constraints by default. However, it is important to note that this may not preserve gate set, connectivity, etc. The optimization involves removing redundancies, applying Clifford simplifications, commuting single-qubit gates to the front of the circuit, and other compiler passes   \cite{Sivarajah_2021}.

Finding the optimal balance between reducing the Trotter error by increasing N$_{\text{trotter}}$ and minimizing the circuit depth is crucial. In this study, we use $dt = 0.05T$. We find that, by inspecting the corresponding matrices, the Trotter error is still acceptable for the system if we use $\Delta t = dt$ (i.e. N$_{\text{trotter}}=1$).

% provides the best results demonstrated later in this paper in Sec.\ref{sec:results}. 

The number of 2-qubit gates in one Trotter step compiled for different architectures is shown in Table \ref{tab:one_Trotter_step}. Note that the circuit parameters are susceptible to the exact optimization strategy and the backend of choice. Moreover, for each time step, the circuit is composed of several sub-circuits and will be re-compiled separately. Due to the re-compilation, the number of 2-qubit gates does not necessarily increase linearly with the number of appended Trotter steps.

\begin{table}[h!]
    \centering
    \begin{tabular}{|c|c|c|c|}
    \hline
   & Uncompiled & IBM & H1 \\
      \hline
   Depth & 529 & 373 & 199 \\  
   Gates, total & 1862 & 596 & 298 \\
   2-qubit gates & 216 & 245 & 120 \\
\hline
    \end{tabular}
    \caption{Circuit parameters per one Trotter step. The circuits are compiled for the corresponding gate sets native to the architecture, i.e. superconducting (IBM) or trapped ion (H1). The optimization level in TKET  is 2  \cite{Sivarajah_2021}.}
    \label{tab:one_Trotter_step}
\end{table}

\subsection{Observables}
\label{subsec:observables}

In this section, we will examine the physical observables (corresponding to what is measured) which reflect the phase transition in the coupled cavities. It is important to note that these observables are time-dependent, so measurements must be taken at each time step of the propagation.

\begin{itemize}[leftmargin=0. in]
\item[]   
\begin{center}
    \textbf {(a) Overlap}
\end{center} 
  
The first observable we will consider is $\Lambda(t)$, which measures the overlap between the initial and the time-propagated wave function at time $t$: 
\begin{align}
\label{eq:lambda}
    \Lambda (t) = -\frac{1}{L} \log_2(|\braket{\Psi_{t=0}| \Psi(t)}|^2)
\end{align}

Equation \ref{eq:lambda} shows that $\Lambda$ is zero for the initial state and reaches a maximum when the overlap between the states is minimal. This means that $\Lambda$ is larger when the system is closer to the superfluid state.

\item[] 
\begin{center}
    \textbf{(b) Total number of excitations and the order parameter}
\end{center} 

The system is characterized by the number preserving polaritonic excitations $N_{\text{ex}}$, which is the sum of atoms in the excited state and photons in each cavity, 
\begin{align}
    N_{\text{ex}} = \sum_i n_i 
 \label{eq:n_ex}   
\end{align}
 with
\begin{align}
\label{eq:ni}
    n_i = (b^\dagger_i b_i + \sigma_{ee}^i)
\end{align}

\noindent where $\sigma_{ee}^i$ is the number of atomic excitations in $i^{th}$ cavity.  The total number of excitations is a conserved quantity and its operator commutes with the Hamiltonian. In terms of $Z$-Pauli gates, $Z_j$, acting on $j^{th}$ qubit,
\begin{align}
\label{eq:n_exq}
    N_{\text{ex}}^{\text{qubit}} &= n_1 + n_2  = \\ \nonumber
    &=  (2 - 0.5 Z_0 - Z_1 - 0.5 Z_2) + \\ \nonumber
    &+ (2 - 0.5 Z_3 - Z_4 - 0.5 Z_5)
\end{align}

The following observable is not conserved:
\begin{align}
\label{eq:n_ex2q}
&Q_{ex}^{ \text{ qubit}} = n_1^2 + n_2^2  =  \\ \nonumber
&= (5.5 - 2Z_0 - 4Z_1 - 2Z_2 + Z_0Z_1 + Z_1Z_2 + 0.5Z_0Z_2) + \\ \nonumber
&+ (5.5 - 2Z_3 - 4Z_4 - 2Z_5 + Z_3Z_4 + Z_4Z_5 + 0.5Z_3Z_5) 
\end{align}

The total excitation variance is also a time-dependent quantity:
\begin{align}
\label{eq:variance}
    \delta N^2(t) = \sum_i^{L=2} \big[\braket{n_i^2} - \braket{n_i}^2 ]
\end{align} 

The order parameter, denoted by ``OP'' and defined by 
\begin{align}
\label{eq:op}
    OP = \frac{1}{T}  \int_0^T \delta N^2(\tau) d\tau,
\end{align}
reflects the mean variance of the total number of excitations $N_{\text{ex}}$. The system becomes increasingly disordered as it moves farther away from the Mott insulator state. In the superfluid state, each cavity can hold up to 2 photons, which means that the cavity array is not ``ordered''. Note that since the two cavities are identical, it is not necessary to measure the polaritonic excitation variance for each of them. 
Note that it is sufficient to measure $\braket{n_i}(t)$ in one cavity and multiply the result by the factor of 2. We have verified this conclusion with statevector simulations. This observation is especially important for hardware experiments meaning fewer measurements are required.

\end{itemize}

\section{Results}
\label{sec:results}

In this section, we show the simulation results for two coupled cavities (see in Sec. \ref{sec:qed}) performed with the Holstein-Primakoff transformation for the multiphoton regime (see Sec. \ref{sec:holstein-primakoff}).
First, the quantum statevector and measurements-based results are compared with the classical simulation to demonstrate the validity of the approach as well as its efficiency. Then we analyze the effect of noise.

\subsection{Noiseless backend results}

\begin{itemize}[leftmargin=0.in]
  \item[] 
\begin{center}
   \textbf{(a) Overlap} 
\end{center}   

\begin{figure}[h!]

\begin{subfigure}[b]{1\textwidth}
\centering
   \caption{Classical and statevector results}
   \vspace{-0.15cm}
   \includegraphics[width=10cm]{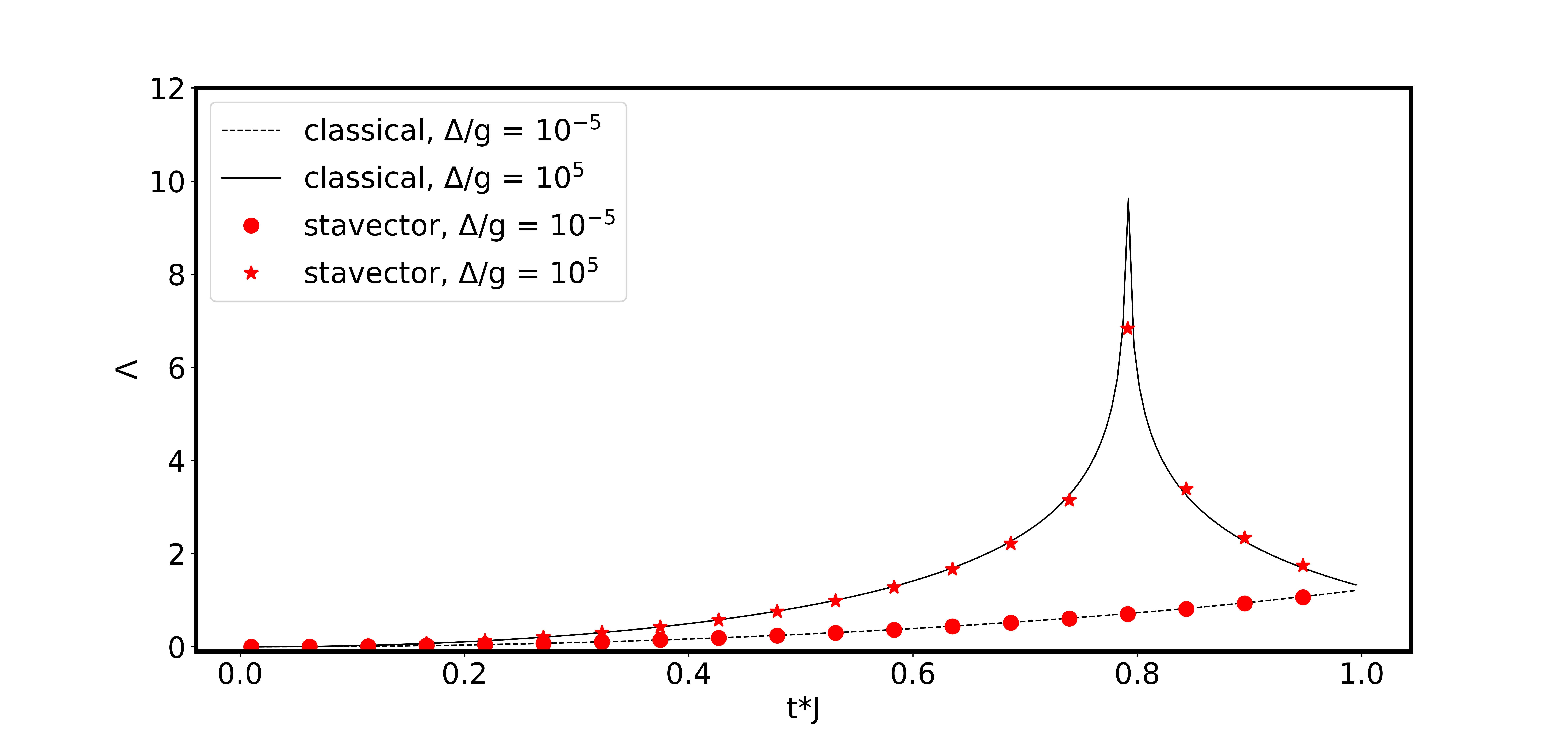}
   \label{fig:lambda1} 
\end{subfigure}

\begin{subfigure}[b]{1\textwidth}
\centering
   \caption{Canonical Swap Test noiseless results}
   \vspace{-0.15cm}
   \includegraphics[width=10cm]{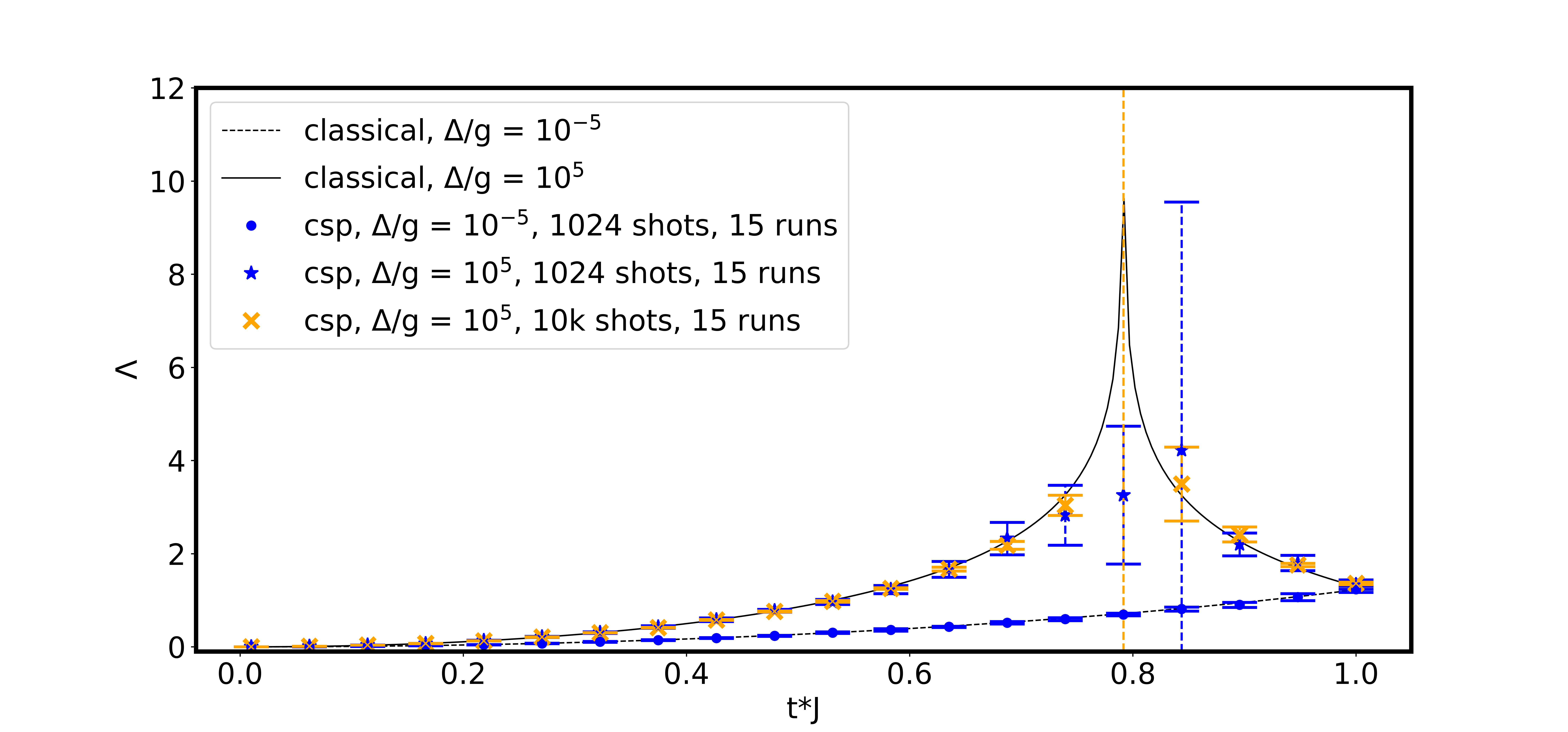}
   \label{fig:lambda2}
\end{subfigure}

\begin{subfigure}[b]{1\textwidth}
\centering
   \caption{Vacuum Test, noiseless results}
   \vspace{-0.15cm}
   \includegraphics[width=10cm]{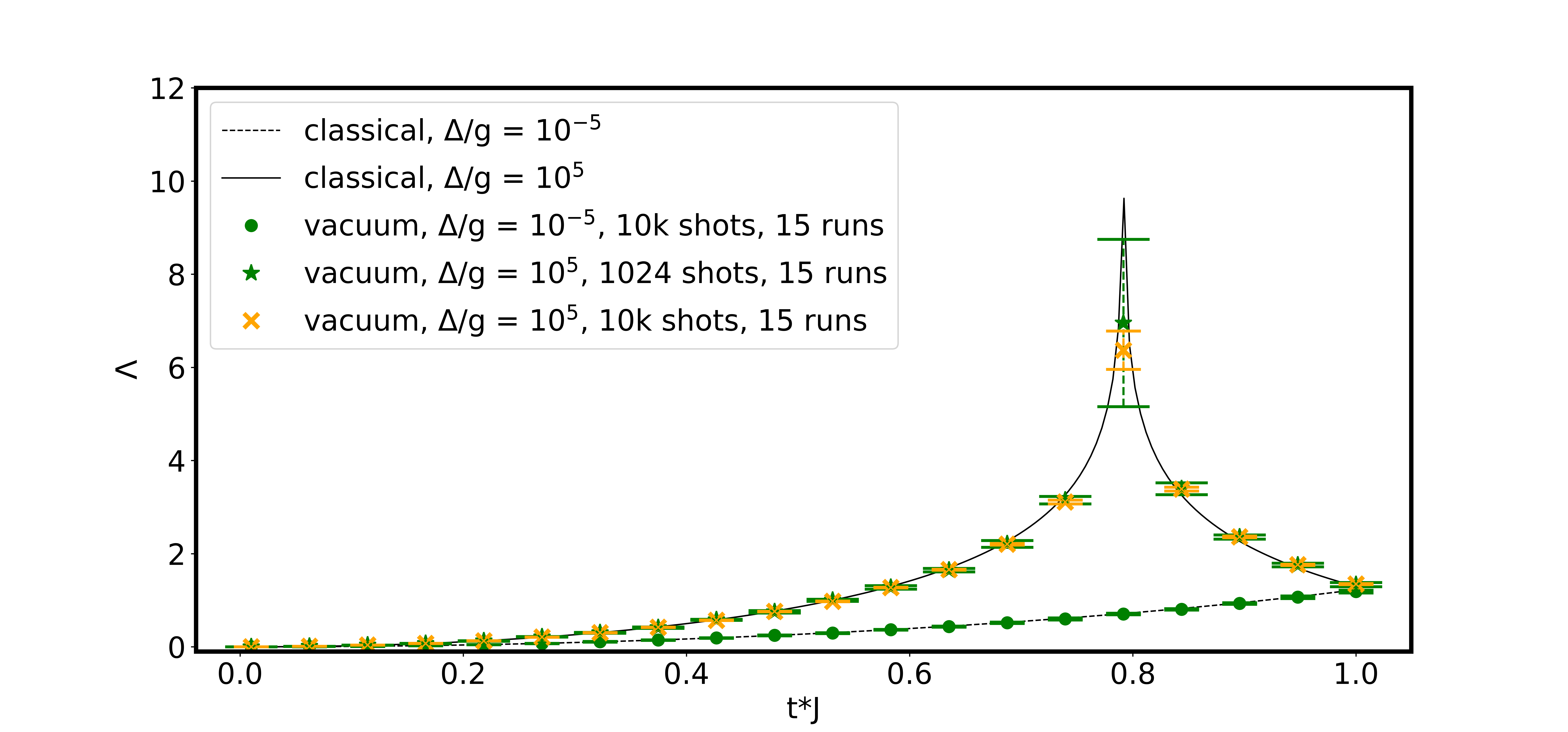}
   \label{fig:lambda3}
\end{subfigure}
\caption{The overlap $\Lambda(t)$ (see Eq. \ref{eq:lambda}) from (a) statevector, 
(b) Canonical Swap Test Protocol   (CSP), (c) Vacuum Test simulations.  This parameter reflects the phase transition and is related to the overlap between the initial Mott-insulator state and the time-evolved wavefunction. 
For the corresponding $\Delta$ values, see the labels in the respective legend boxes. 
The data correspond to the mean value over $N_{\text{runs}}=15$. The error bar shows the confidence interval for the 99$\%$ confidence level with $Z$-factor equal to 2.576 (see Eq. \ref{eq:confidence_interval}).}
\label{fig:lambdas}
\end{figure}  

We start with examining the phase transition by analyzing $\Lambda(t)$, which is a function of the overlap of the initial and the evolved state. $\Lambda(t)$ is given by Eq. \ref{eq:lambda} in Sec. \ref{subsec:observables} (a). Fig. \ref{fig:lambda1} displays the classical and statevector results for $\Lambda(t)$. The black curves (no markers, solid upper line and dashed lower line) represent the classical results as described in detail in   \cite{Norambuena2020}. In this approach, each operator, including the Trotter operator, is represented with its corresponding matrix, and the dynamical simulation is carried out by matrix multiplication. For better accuracy, the time step in the classical calculation is ten times smaller than the time step we used in the subsequent quantum calculations. We have verified that if $\Delta, \: g, \: \omega_0, \: J $ are similar to those in   \cite{Norambuena2020}, then our results are identical to theirs. 

Each classical curve in Fig. \ref{fig:lambda1} corresponds to a different detuning value $\Delta/g$ = $\{ 10^{-5}, 10^5\}$, with $g$ being fixed (refer to Table \ref{tab:params}). When the detuning is negligible, $\Lambda(t)$ remains small and gradually increases with time corresponding to a  small-scale ``leakage'' of the Mott insulator wavefunction. As the detuning becomes more significant, the system undergoes a phase transition. For $\Delta/g = 10^5$, a sharp, nearly delta function-like peak appears around 0.8$1/J$. The peak position shifts slightly depending on the detuning value (not shown).

The data points shown in red in  Fig. \ref{fig:lambda1} correspond to the quantum statevector simulations for the two values of the detuning. The statevector 
results (red markers) very closely match the classical ones. The small discrepancy around the peak at $t*J \approx 0.8$ at $\Delta/g = 10^5$ can be attributed to the Trotter error due to the larger time step in the statevector simulations, which affects the accuracy when reproducing the cusp. 

\begin{figure}[h!]
    \centering
    \begin{subfigure}[t]{0.25\textwidth}
        \includegraphics[height=1.9cm]{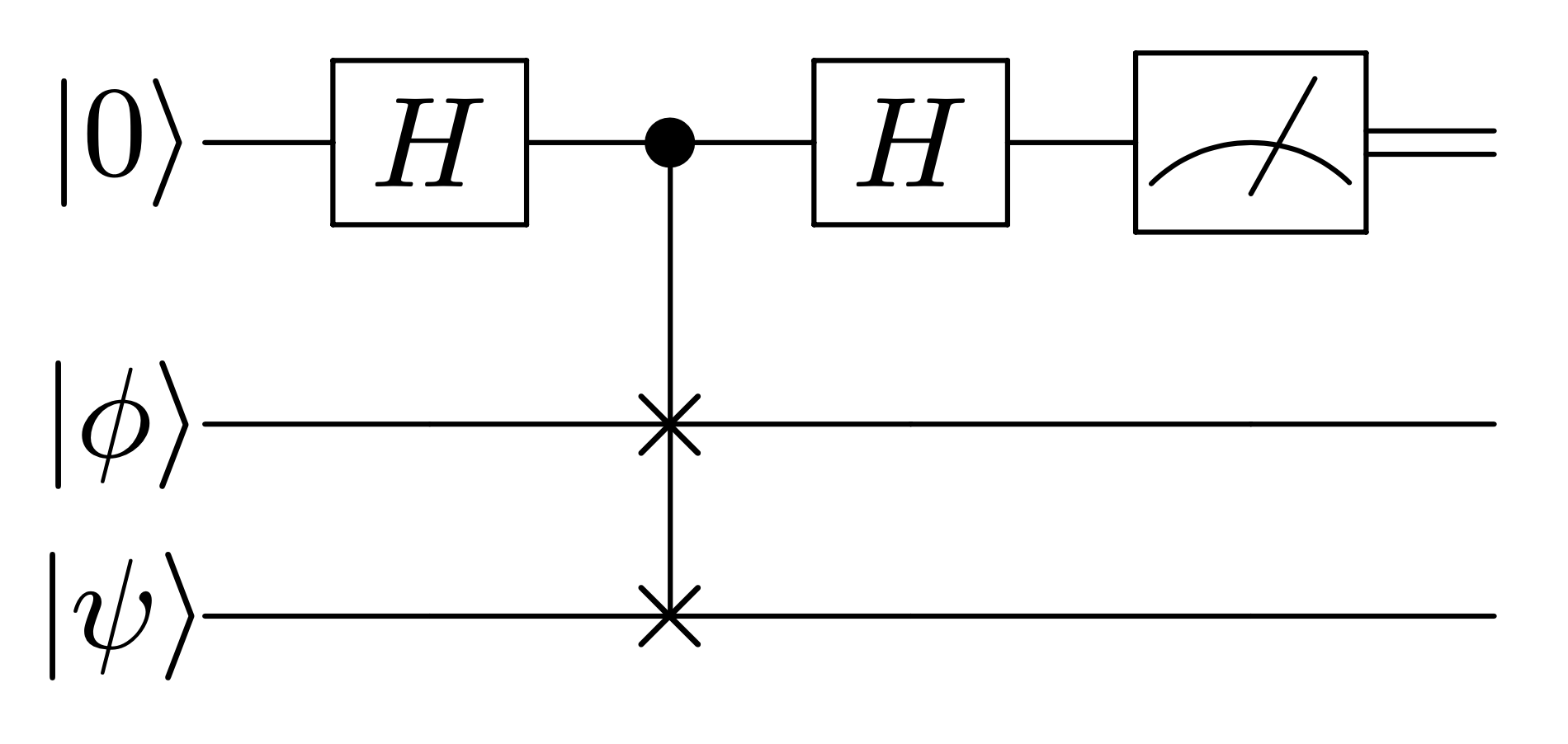}
        \caption{Canonical Swap Test scheme}
        \label{fig:csp}
    \end{subfigure}%
    ~ 
    \begin{subfigure}[t]{0.25\textwidth}
        \centering
        \includegraphics[height=2.5cm]{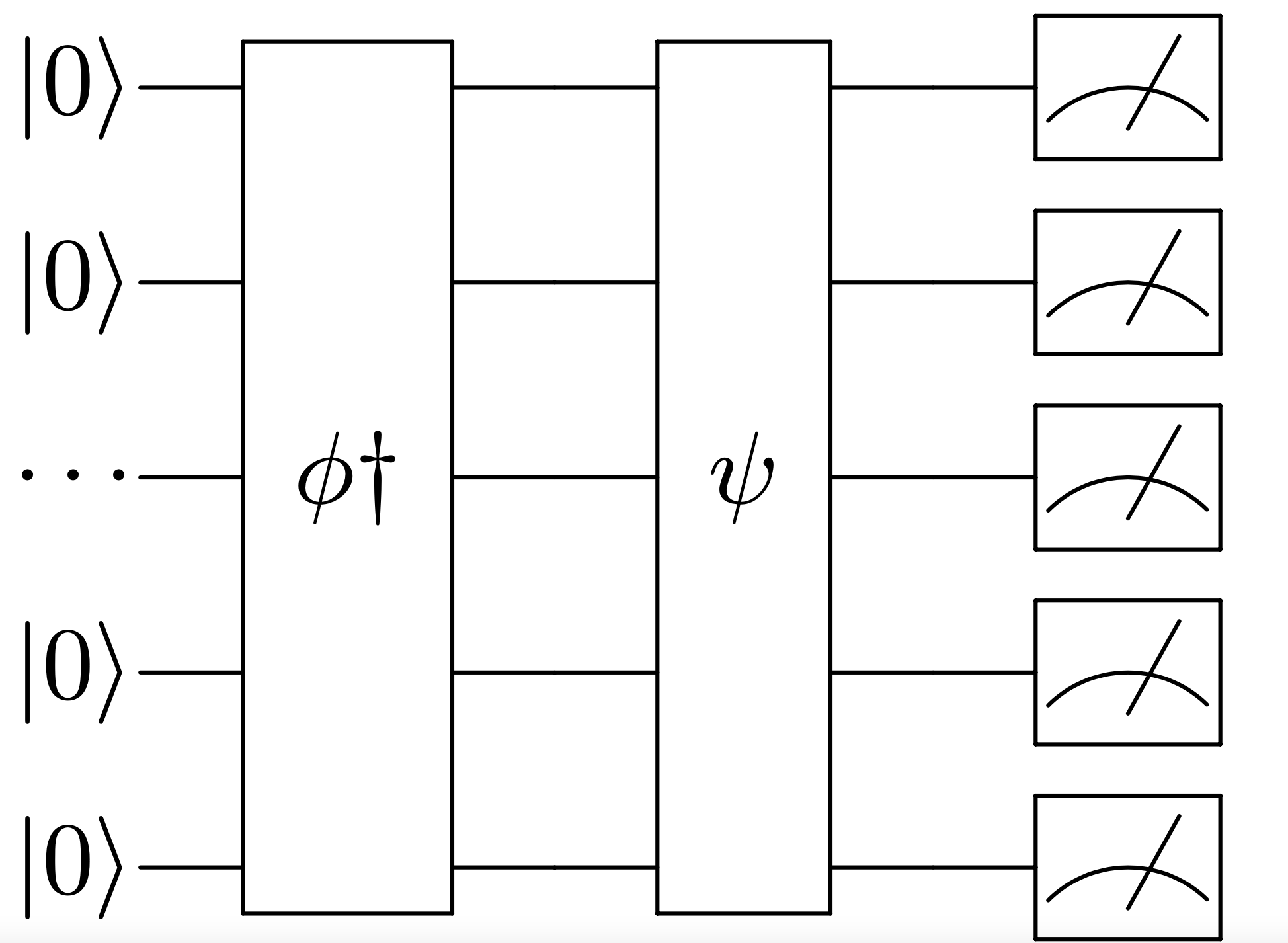}
        \caption{Vacuum Test scheme}
        \label{fig:vacuum}
    \end{subfigure}
    \caption{Schematic representation of the overlap measurement protocols used for $\Lambda$ from Eq. \ref{eq:lambda}}
    \label{fig:overlaps}
\end{figure}

After confirming the accuracy of our mapping scheme in generating statevector results, we move on to shot-based quantum calculations of $\Lambda(t)$. Measuring the overlap between the initial state $\ket{\phi}$ and the propagated wavefunction $\ket{\psi(t)}$ can be done using various measurement protocols, and we compare the ancilla-based Canonical Swap Test Protocol (CSP)   \cite{Barenco97} to the protocol we call Vacuum Test  \cite{Lee_2018}.  
We ran each setup $N_{\text{runs}} = 15$ times to make sure enough data was collected to conclude about the applicability of the protocols. The mean value of $\Lambda$ and the error bars showing the confidence interval, $CI$, are presented in Fig. \ref{fig:lambdas}.  The standard deviation, $\sigma_\Lambda$, is calculated based on the standard deviation of the measured overlap squared. The confidence interval corresponds to the confidence level $99\%$. For each time step,
\begin{align}
\label{eq:confidence_interval}
    P=|\braket{\Psi_{t=0}| \Psi(t)}|^2 \\
    \nonumber 
    \sigma_{\Lambda} = \sqrt{(\frac{d\Lambda}{dP} \sigma_P)^2} \\
    \nonumber
    CI = 2.576*\sigma_\Lambda/\sqrt{N_{\text{runs}}}
\end{align}

The circuit for CSP is shown schematically in Fig. \ref{fig:csp}. The test involves constructing a register with two sub-registers corresponding to the overlapping wavefunctions, and the result is measured on the ancillary qubit similar to the Hadamard test. The probability of finding the ancillary qubit in the $\ket{0}$ state is then $1/2 + 1/2|\braket{\phi|\psi}|^2$. The circuit width is twice the number of qubits in the state plus one. The multicontrol gate shown in Fig. \ref{fig:csp} also leads to increased depth. 
Notably, since the measurement is performed on the ancillary qubit only, with a modest number of shots and when the true value of $P=|\braket{\Psi_{t=0}| \Psi(t)}|^2$ is close to zero, statistical deviations may lead to a negative result in a particular run ($\ket{0}$ is measured slightly less often than $\ket{1}$). This result is nonphysical, however, it cannot be discarded from the statistics since otherwise a bias would be introduced. Instead, many runs are required to achieve the true near-zero mean for $P$. At the same time, for the almost-zero overlap but non-zero standard deviation $\sigma_P$, the confidence interval may become large since $\frac{d\Lambda}{dt}$ is also large (see Eq. \ref{eq:confidence_interval}).  

The results of shot-based CSP simulation are shown in Fig. \ref{fig:lambda2}. 
At the detuning $\Delta/g = 10^{-5}$ (blue circles along the lower classical curve), the wavefunction closely resembles the Mott insulator and changes smoothly on the entire time scale leading to a smooth monotonically increasing $\Lambda(t)$. In this case, the process is near-resonant, and light-matter interaction within each cell dominates the process. CSP measurement result of the observable with 1024 shots is thus very accurate nearly matching the classical data. The confidence interval does not exceed 0.1. However, the accuracy is worse for very large detuning. In particular, there is a discrepancy in the region around the sharp maximum of $\Lambda(t)$ when the overlap between the initial and evolved wavefunctions is small. Blue stars overlaying the upper classical curve show the mean result of the measurement with 1024 shots. The peak position is not reproduced well. The error bar around the peak is very large. The quality is increased when runs are performed with 10000 shots (orange crosses close to the upper classical curve). At the peak, the overlap is found to be almost zero leading to large values of $\Lambda$. However, the error around the peak remains large as well.

Next, we employ the Vacuum test (see Fig. \ref{fig:vacuum}), which works by appending the circuits corresponding to $\bra{\phi}$ and $\ket{\psi}$ and measuring the result directly on every qubit. This eliminates the need to duplicate the number of system qubits and add an ancillary qubit, but increases the depth of the circuit. The width of the circuit is that of the wavefunction. The result is equal to the probability of measuring the all-zero state on all qubits. The corresponding $\Lambda$ is shown in Fig. \ref{fig:lambda3}, where the green circles (bottom) and stars (top) correspond to the noiseless result for $\Delta/g = 10^{-5}$ and $10^5$, respectively. Orange crosses closely following the upper (classical) curve show the result for the large detuning and large number of shots.
Again, in the case of small detuning, the simulation results lie directly on top of the classical ones well within the confidence interval of 0.1. For the large detuning, the results are also in good agreement with the classical prediction, except for the peak region, where the obtained values are very large and lie outside the plot range. Similarly to what is described above, this would correspond to the situation when the measured overlap is very close to zero, so $\Lambda(t)$ approaches infinity. We add a vertical dashed line to indicate this peak. The statistical error is small for both small and large detuning everywhere except the peak point.

% We also tested whether the results could withstand adding a small readout error, represented by yellow stars in Fig. \ref{fig:lambda3}. We found that the shape of $\Lambda$ was reproduced, and the results in the peak region also resembled a delta function. Therefore, the Vacuum Test protocol allows us to reproduce the classical results well enough to hope a noisy simulation or a hardware run can help detect the phase transition.

Comparing Fig. \ref{fig:lambda2} and Fig. \ref{fig:lambda3} shows that the Vacuum Test is more suitable for the cavity QED problem under consideration. Using CSP protocol overlooks the singularity around the peak at large detuning and requires many runs and a large number of shots because the measurement is performed on a single qubit. At the same time, smoothing the peak makes the CSP protocol less suitable for studying the phase transition. The Vacuum Test, on the contrary, amplifies the transition. Importantly, the Vacuum Test performs better with a modest number of shots since its results nearly exactly match the classical ones at all times before and after the cusp. With the number of shots as large as 10000, the Vacuum Test still outperforms CSP in terms of accuracy to the left and the right of the cusp.

% \item[(b)] \textbf{Total number of excitations and the order parameter}
  \item[] 
\begin{center}
   \textbf{(b) Total number of excitations and the order parameter} 
\end{center}   

We now turn to calculating the order parameter, $OP$, from Eq. \ref{eq:op}. This parameter is obtained by finding the $mean \: value$ of the variance after Trotterized time propagation. The result in Fig. \ref{fig:ops} is plotted as a function of detuning (rather than time). 

Before presenting our results, let us discuss and interpret the classical simulation from   \cite{Figueroa2018} (Fig. \ref{fig:ops}, inset). In that paper, the initial order parameter is close to zero, indicating that the system is in the Mott-insulator state, and adding more photons to each cell is prohibited. As the detuning becomes larger, the phase transition occurs. The resulting curve for the order parameter consists of two distinct plateaus. In contrast to that study, we consider a much stronger coupling for reasons mentioned in Section \ref{subsec:system_params}. Under the conditions described in Table \ref{tab:params}, the classical curve for the order parameter is shown as the black solid line in Fig. \ref{fig:ops}. Just like the result in  \cite{Figueroa2018}, it features two plateaus which are connected by a smooth transition around $\Lambda/g = \mathcal{O}(1)$. The initial order parameter is around 0.6 at small detuning before the phase transition. 
The nonzero value is due to the variance not being consistently small throughout the characteristic period in the strong-coupling regime.

\begin{figure*}
     \begin{subfigure}[b]{1\textwidth}
     \centering
         \includegraphics[width=10cm]{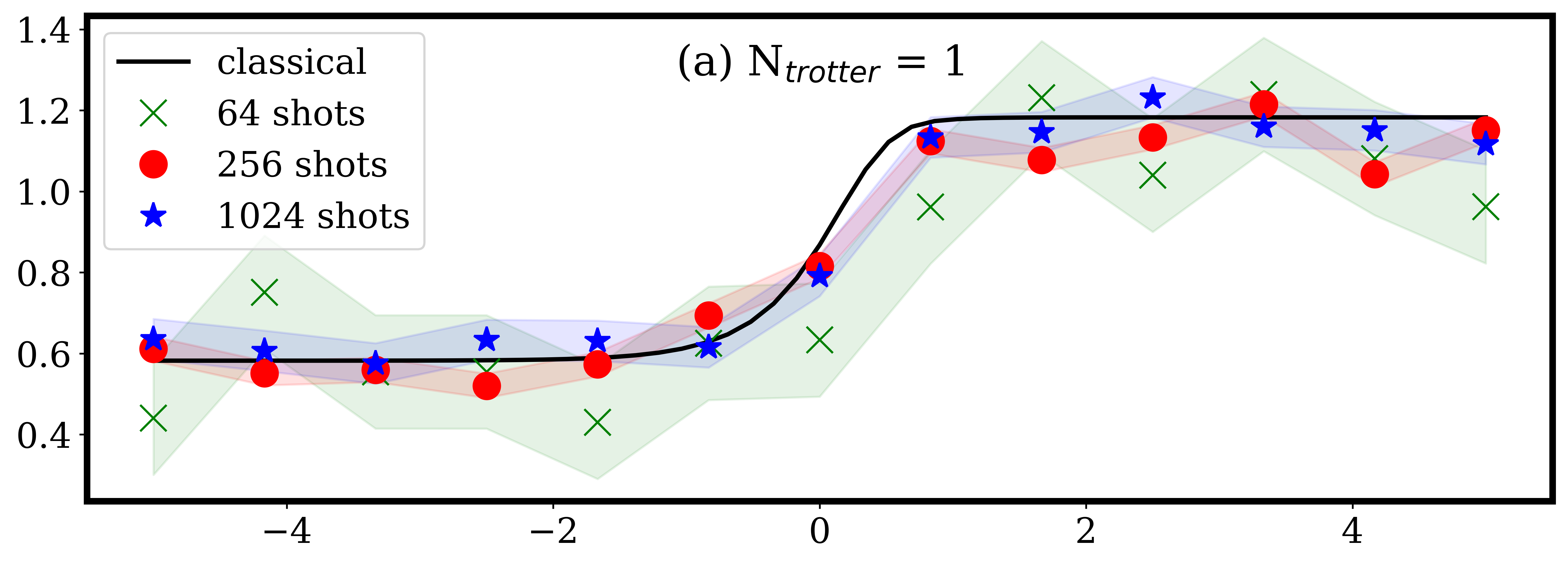}
        \label{fig:ops_a} 
     \end{subfigure}
     \begin{subfigure}[b]{1\textwidth}
     \centering
     \vspace{-0.44cm}
     \hspace{-0.5cm}
         \includegraphics[width=10.4cm]{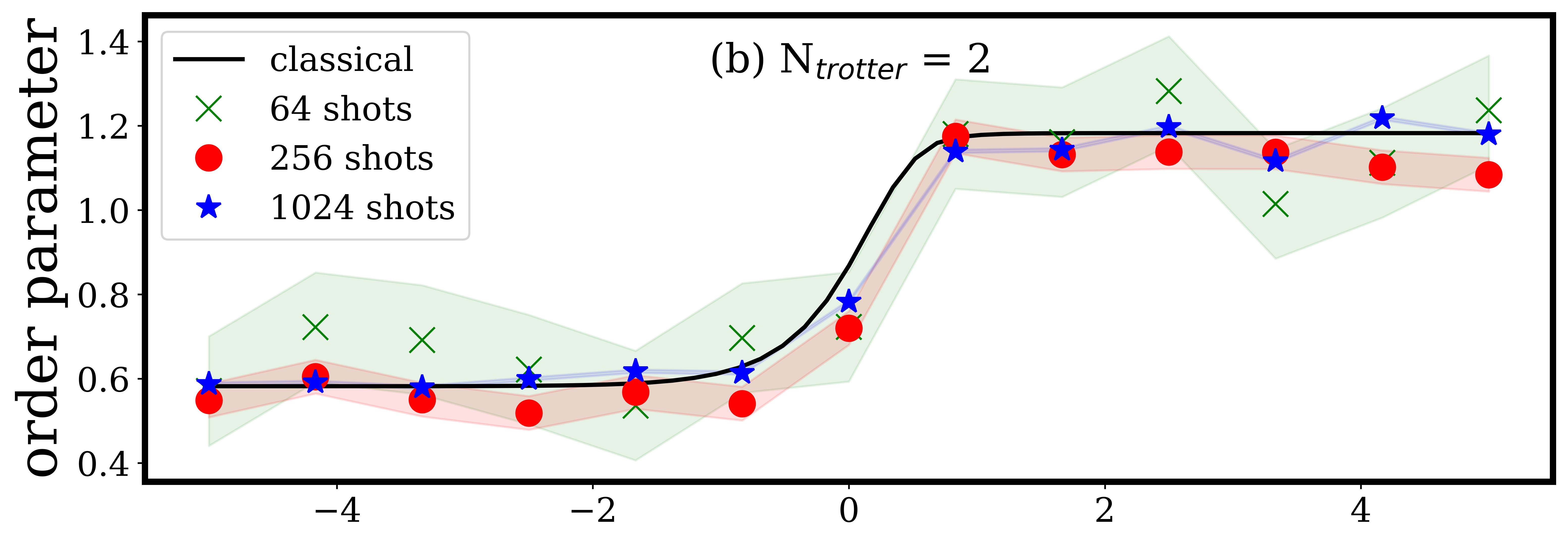}
   \label{fig:ops_b}
     \end{subfigure}
     \begin{subfigure}[b]{1\textwidth}
     \centering
     \vspace{-0.42cm}
         \includegraphics[width=10cm]{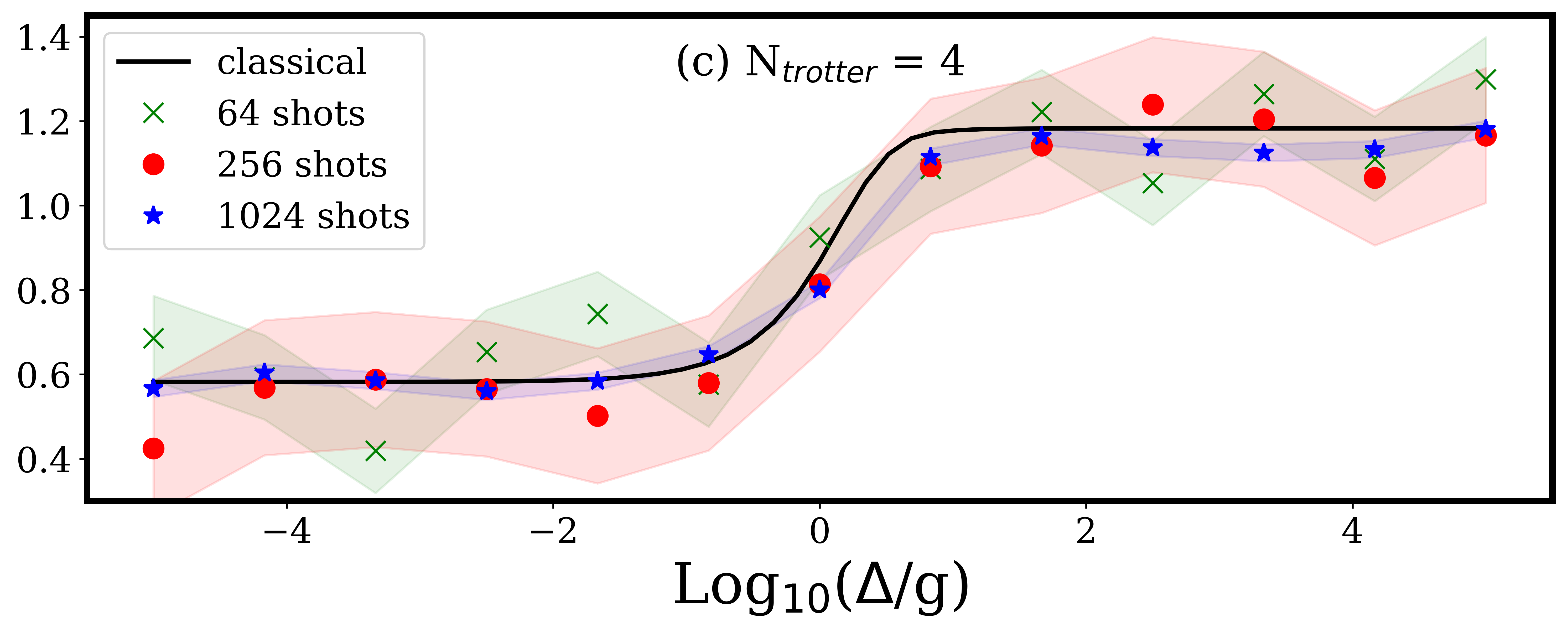}
   \label{fig:ops_c}
     \end{subfigure}
 \vspace{0.cm}    
\caption{Order parameter as a function of the detuning. Black curve: classical simulations. Dots and stars:  noiseless shot-based simulations with various numbers of Trotter steps per time step and different number of shots. Transparent shadows correspond to the mean difference with the classical result for each curve with the corresponding color. Inset: a picture adopted from   \cite{Figueroa2018} to show the comparison with the QED classical study. N$_{\text{trotter}}$ is the number of Trotter steps per time step. }
\label{fig:ops}
\end{figure*}

We investigate the effect of varying the number of Trotter steps (within just one time step) and the number of shots on the results of a noiseless simulation. The number of time steps remains constant at 20. 
Fig. \ref{fig:ops} has three series of results: 64 shots (green, crosses), 256 shots (red, circles), and 1024 shots (blue, stars), each with a different number (1-4) of Trotter steps per time step. The figure shows that, in the given range of Trotter step sizes and shot numbers, increasing the number of shots has a greater impact on result precision compared to reducing the Trotter step size. Another point is that since the cavities are identical, measuring one cavity with $X$ shots is equivalent to measuring $L$ cavities with $X/L$ shots each.
The best result is represented by the blue dots with the mean error of only $\epsilon = 0.02$. Importantly, these results show that the phase transition can be reproduced with a modest number of shots. This is because the required level of accuracy in this problem is different from that required in typical quantum chemistry simulations. Instead of seeking to find the exact value of a parameter, we aim to determine the approximate value of the detuning at which the phase transition occurs.

\end{itemize}

\subsection{Noisy backend results}
\label{section:noisy}

 To assess whether current quantum computers can handle the problem described in this paper, we perform emulator-based simulations using two different architectures. For the superconducting device, we use the Aer backend with a noise model that matches that of the IBMQ Montreal machine   \cite{Qiskit}. Meanwhile, we used Quantinuum's H1-2E emulator to simulate an ion-trap device. In general, superconducting devices offer faster simulations but with more noise, while ion-trap machines provide lower noise but slower simulations.

\begin{figure}
     \begin{subfigure}[b]{0.5\textwidth}
    \caption{}
    \vspace{-0.15cm}
         \includegraphics[width=7.cm]{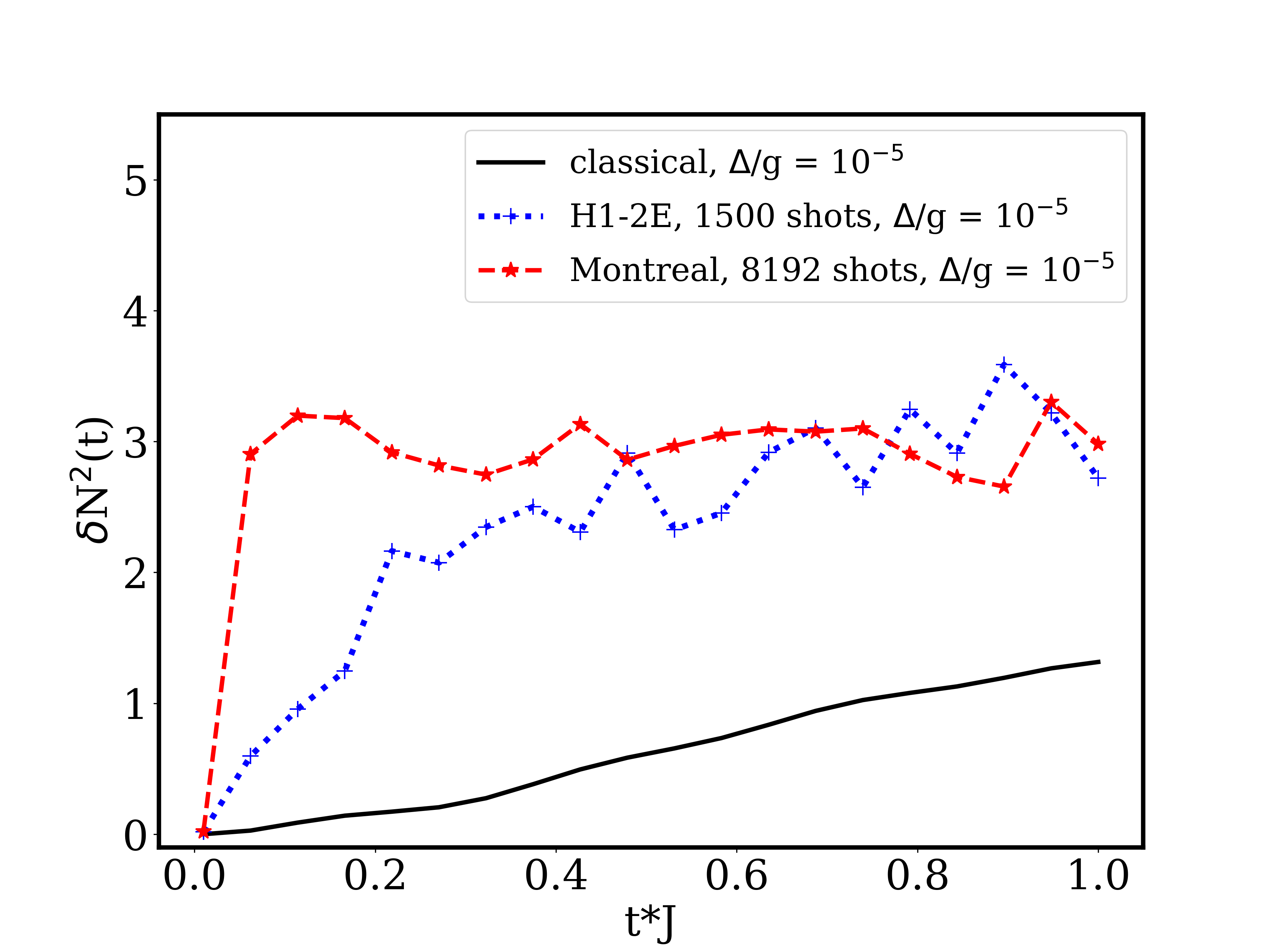}
        \label{fig:noisy_e-5} 
     \end{subfigure}
     \begin{subfigure}[b]{0.5\textwidth}
    \caption{}
     \vspace{-0.15cm}
         \includegraphics[width=7.cm]{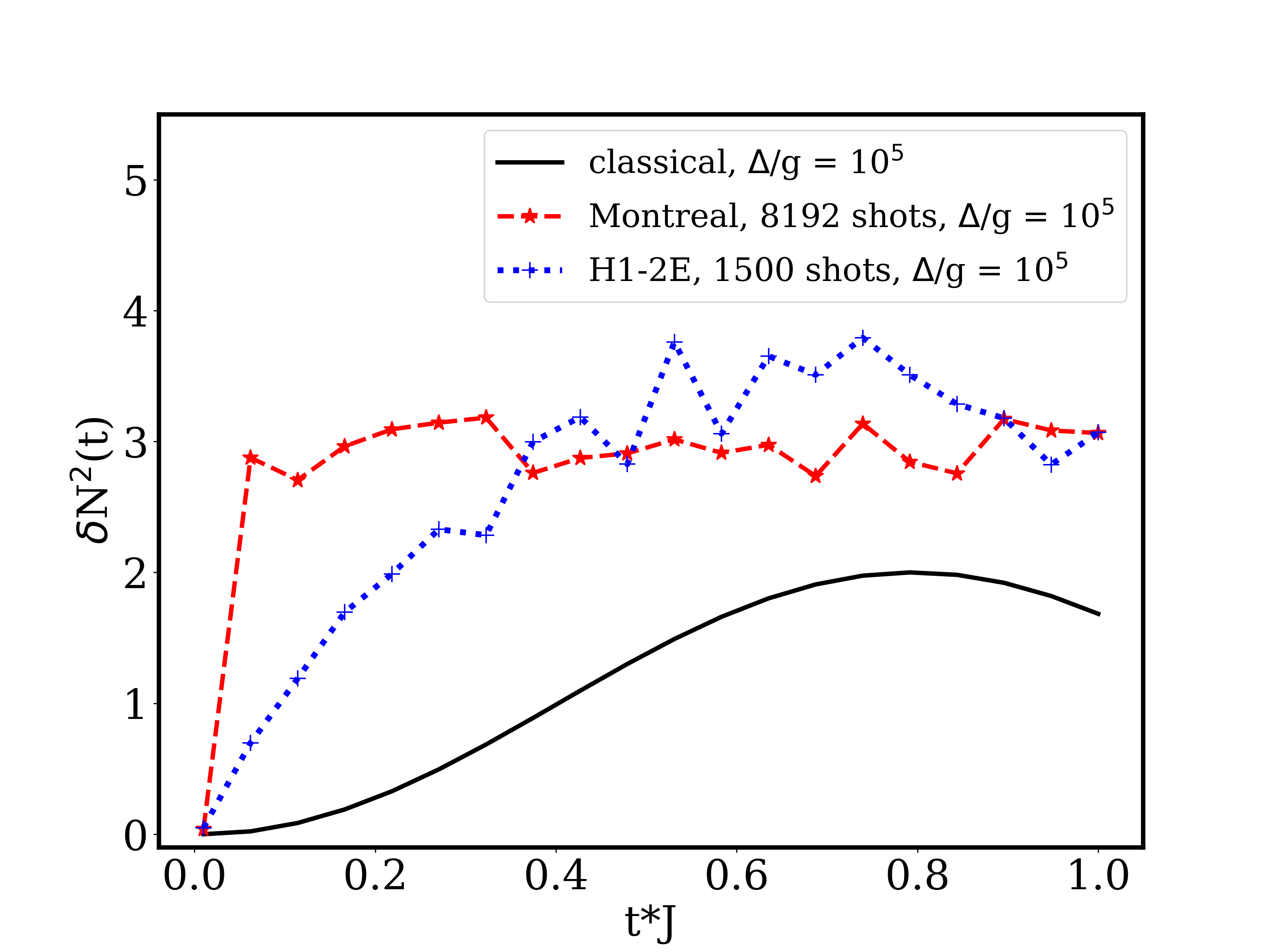}
   \label{fig:noisy_e5}
     \end{subfigure}
\caption{Time-dependent variance in the number of excitations, 
$\delta N^2(t)$, see Eq. \ref{eq:variance}, \ref{eq:n_ex},  \ref{eq:n_exq}, \ref{eq:n_ex2q}. 
(a) Small detuning, $\Delta$/g = 10$^{-5}$. (b) Large detuning, $\Delta$/g = 10$^{5}$.
}
\label{fig:noisy}
\end{figure}

Although the final goal is to observe the phase transition such as in Fig. \ref{fig:ops}, the order parameter for each value of the detuning is calculated as a mean over time, and therefore it does not allow for assessing the performance of the dynamics algorithm in detail.
Instead of reproducing the order parameter graph with a noisy simulation, Fig. \ref{fig:noisy} shows the curves for the variance in the number of excitations, $\delta N^2$, Eq. \ref{eq:variance} as a function of time, for two extreme values of the detuning, $\Delta/g = \{10^{-5}, 10^5\}$, Fig. \ref{fig:noisy}. 
When the detuning is negligible, we expect the curve to look like the black curve in \ref{fig:noisy_e-5}, i.e. to monotonously increase. For very large detuning, the classical result in Fig. \ref{fig:noisy_e5} has a maximum at $t \approx 0.8/J$. 
The order parameter, as a mean of $\delta N^2(t)$ will, therefore, be sensitive to the time-dependent behavior of the number of excitations. To claim the effect has been observed in an experiment in the entire detuning range, one needs to detect a transition between two plateaus (see Fig. \ref{fig:ops}). If one observes that a noisy result is shifted up from the classical result due to some coherent noise - but that the shape of the curve is reproduced - we can expect the phase transition to be detected. Noise mitigation techniques such as, for example, zero-noise extrapolation may potentially improve the result.

The results obtained from an ion-trap emulator are shown in Fig. \ref{fig:noisy}. Although the absolute value of the variance becomes significantly higher than expected after just a few Trotter steps, the overall shape of the curve is reproduced. Due to limited computational resources, we do not build the full curve (such as in Fig. \ref{fig:ops}) for the order parameter before attempting to mitigate the noisy run. 

Yet, emulating the process on Aer backend with a realistic noise model (see Fig. \ref{fig:noisy_e-5}, \ref{fig:noisy_e5}) produces qualitatively similar results for both small (top panel) and large (bottom panel) detuning. Both start at some small values and quickly reach a plateau of $\delta N^2 \approx 3$. One can see why this is happening if we look at Eqs. \ref{eq:n_ex}, \ref{eq:n_exq}, \ref{eq:n_ex2q}, and \ref{eq:variance}.
If we assume very large noise, all the results for nonconstant operators average out and only the constant value survives, $\big(\braket{n_1^2} - \braket{n_1}^2 + \braket{n_2^2} - \braket{n_2}^2)_{\text{noise} \rightarrow \infty} \rightarrow 1.5 +1.5 = 3$, as we indeed see in the figure.

\subsection{Scaling}
\label{section:scaling}

\begin{figure}
     \centering
     \begin{subfigure}[b]{0.45\textwidth}
    \caption{}
         \hspace{-0.8cm}
         \includegraphics[width=7.6cm]{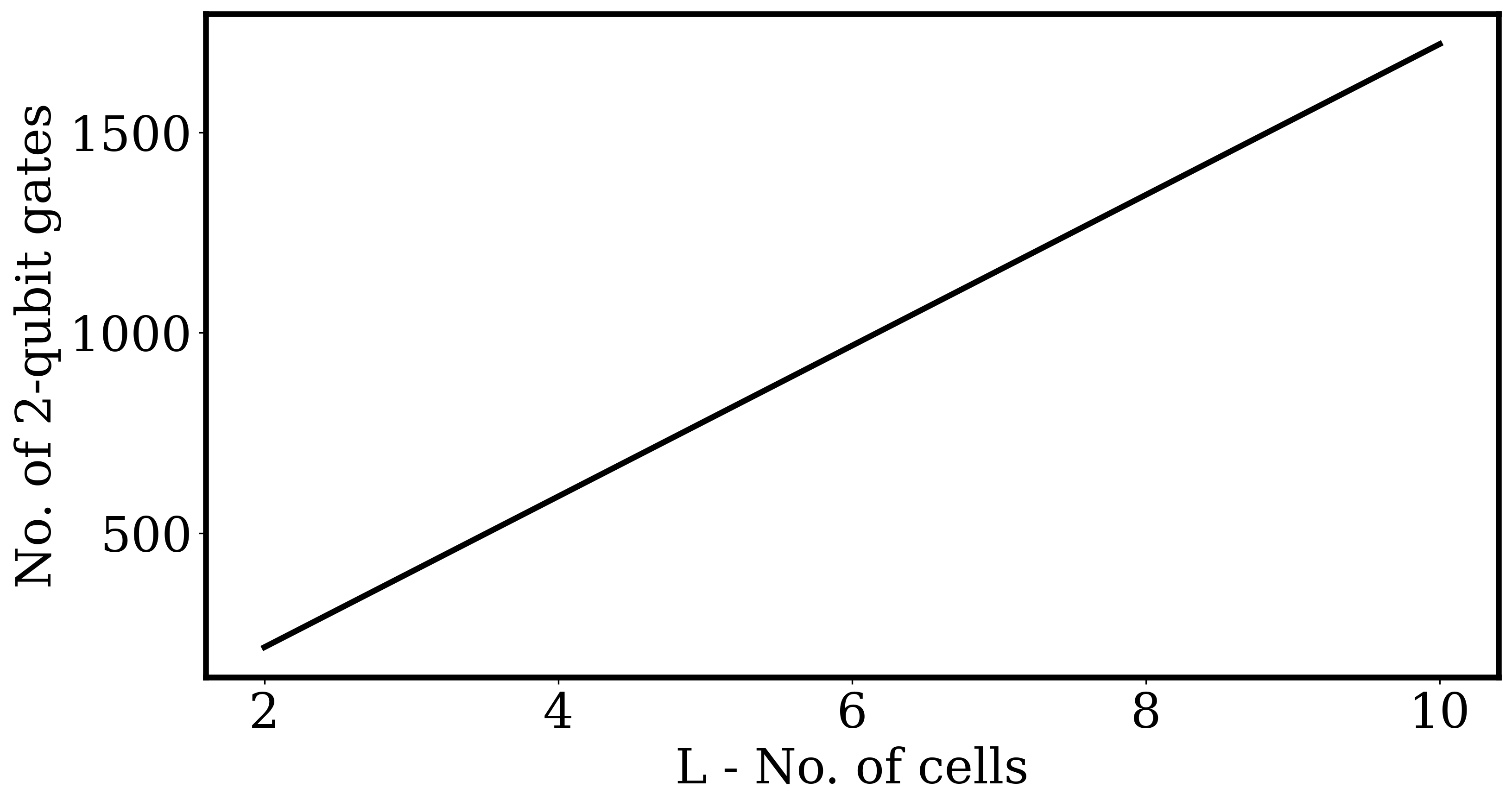}
         \vspace{-0.2cm}
         \label{fig:scaling}
     \end{subfigure}
     \begin{subfigure}[b]{0.45\textwidth}
        \caption{}
         \includegraphics[width=7.3cm]{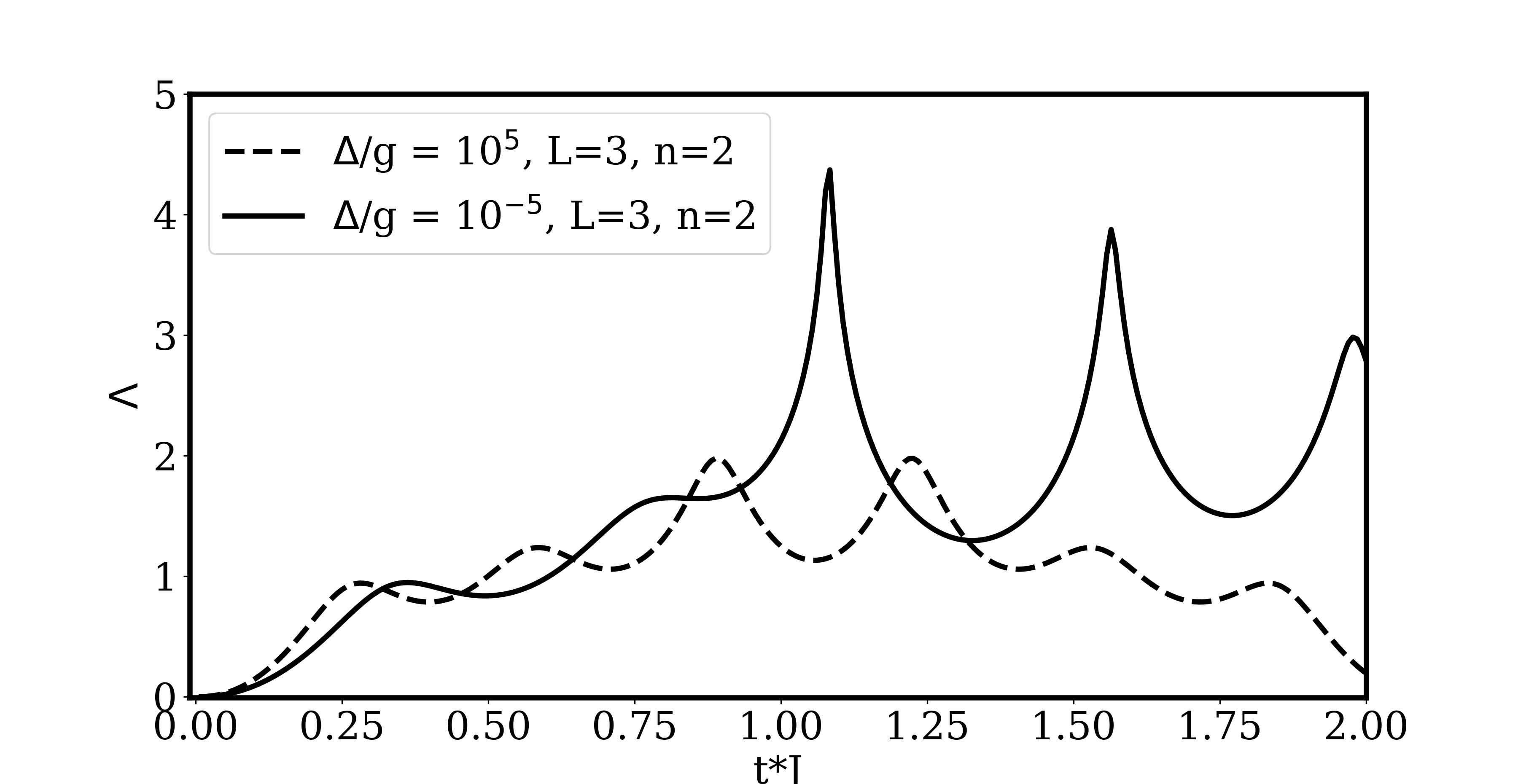}
         \vspace{-0.3cm}
         \label{fig:scaling_1}
     \end{subfigure}
   \caption{(a) Number of 2-qubit gates per one Trotter step. The circuits are generated by TKET without taking an architecture into account, i.e. for $L=2$, the circuit is the same as in the "Uncompiled" column of Table \ref{tab:one_Trotter_step}. L=10 corresponds to a 30-qubit calculation. (b) Classical simulation of $\Lambda(t)$ for  $L=3$, and the number of photons in the band $n=2$, and different detuning.}
\end{figure}

We have analyzed how the circuit parameters will scale for larger systems with the number of cells $L > 2$. Fig. \ref{fig:scaling} shows that the number of 2-qubit gates per uncompiled circuit with one Trotter step changes linearly. Note that one extra cell corresponds to 3 extra qubits, so the scale goes from 6 qubits (as is the focus of this study) to 30 qubits. 

Finally, Fig. \ref{fig:scaling_1} shows classical results for $L=3$ and $n=2$, where $n$ is like in Eq. \ref{eq:cavity_state}, on the time scale beyond $T=1/J$. The characteristic period for different numbers of cavities and higher number of photons changes. For small detuning (the green curve), the result does not appear to be periodic neither it is monotonous for $t<2/J$. Therefore, for a quantum simulation of larger systems not only a bigger register but also a longer simulation time is needed.

\section{Discussion and further steps}
\label{sec:conclusions}
In conclusion, we have introduced a scheme for mapping bosons in a mixed spin-boson system that relies on the inverse Holstein-Primakoff mapping. Our method has been tested in the multiphoton regime and on systems with several atoms described with spin operators. Comparing our results with classical simulations confirms the validity of the proposed mapping. We expect this method to be at least as efficient as the binary mapping. Further, our results with noiseless simulations with only the stochastic error present demonstrate that the phase transition can be detected with a relatively large Trotter step and a modest number of shots which is realizable on near-term quantum hardware. 
% When recording the overlap between the evolved and the initial wave function, the phase transition manifests as a sharp peak in the time dependency of $\Lambda$, eq. \ref{eq:lambda}.

However, upon testing the model on noisy emulators, we find that further improvements to the dynamics simulation algorithm are necessary due to the large number of 2-qubit gates per time step in the standard Trotter-Suzuki scheme, which lead to large amounts of noise introduced in the simulation. The ion trap runs suggest that the use of noise mitigation may lead to more accurate simulation results.
However, the noise level of the superconducting machines is too high for noise mitigation techniques alone to be effective. Thus, we believe that further exploration of other algorithms, such as those relying on randomly compiling a Trotter step circuit  \cite{Campbell2019}, particularly for higher values of the detuning, may lead to significantly improved results in combination with noise mitigation.

For future work, we plan to generalize the method extending to systems with even more photons or higher maximum spin values. For very high spins, it would be beneficial to analyze the error introduced by truncating the Newton series.
With tens of logical qubits available, one may be able to simulate processes involving hundreds to thousands of photons. This will open the door to modeling important collective effects in cavity QED, such as, for example, superradiance.

In regards to the unitary circuit mapping, an alternative approach to Trotter-Suzuki decomposition is using block encoding techniques \cite{Low2019hamiltonian} which prove to be more efficient in the asymptotic limit. These methods can be explored in combination with Holstein-Primakoff mapping with higher spins as well as for boson operators. 

In larger systems, the issue of the number of shots that can be performed realistically may become important. In this case, techniques such as classical shadows  \cite{classical_shadows} can be exploited to reduce the simulation cost. Additionally, we may explore mapping polaritonic excitations, rather than dividing the register into spin and boson components.
This approach can potentially reduce the requirement for the number of qubits.
Once noise mitigation has been tested successfully on an emulator, we may consider running the model on real quantum hardware.
% Expressing the Hamiltonian in terms of polaritonic ladder operators and mapping them to qubits can potentially be an effective strategy and will be considered in future work.

\section{Acknowledgements}
The authors would like to acknowledge their colleagues at Quantinuum, particularly, we are particularly grateful to Michelle Sze for discussions about the cavity QED problem and boson mapping, as well as for general remarks on the paper's text. We also extend our thanks to Ramil Nigmatullin for his very helpful comments, and to Nathan Fitzpatrick, Andrew Tranter, Gabriel Greene-Diniz, and Irfan Khan for assisting with the scientific content. We value the insightful discussions on mapping schemes, circuit compilation and optimization, operator construction, and the Givens rotations ansatz. We appreciate the technical support from Vanya Eccles and John Children in running the emulator, and from Seyon Sivarajah in assisting with TKET-based circuit compilation. 
\newpage
\bibliography{main_arxiv_v2} % Produces the bibliography via BibTeX.

\end{document}